\documentclass[iop]{emulateapj}
\usepackage{color}

\def\aap{{A\&A}}
\def\apjl{{ApJ}}
\def\apjs{{ApJS}}

\def\msol{M$_{\sun}$}
\def\kms{$\frac{\rm km}{\rm s}$\ }
\def\density{$\frac{M_{\sun}}{\rm pc^3}$}
\def\sdensity{$\frac{M_{\sun}}{\rm pc^2}$}

\shorttitle{Counter-rotating disk formation}
\shortauthors{Alig et al.}

\begin{document}

\title{Numerical simulations of the possible origin of the two sub-parsec scale and counter-rotating stellar disks around SgrA*}

\author
{
C. Alig\altaffilmark{1}\altaffilmark{2}, 
M. Schartmann\altaffilmark{1}\altaffilmark{2}, 
A. Burkert\altaffilmark{1}\altaffilmark{2}\altaffilmark{4}, 
K. Dolag\altaffilmark{1}\altaffilmark{3}
}

\altaffiltext{1}{Universit\"ats-Sternwarte M\"unchen, Scheinerstr.1, D-81679 M\"unchen, Germany}
\altaffiltext{2}{Max-Planck-Institut f\"ur extraterrestrische Physik, Postfach 1312, Giessenbachstr., D-85741 Garching, Germany}
\altaffiltext{3}{Max Planck-Institut f\"ur Astrophysik, Karl-Schwarzschild-Str. 1, D-85748 Garching, Germany}
\altaffiltext{4}{Max-Planck-Fellow}


\begin{abstract}

We present a high resolution simulation of an idealized model to
explain the origin of the two young, counter-rotating, sub-parsec scale stellar disks around the
supermassive black hole SgrA* at the Center of the Milky Way. In our model, the collision of a single molecular cloud
with a circum-nuclear gas disk (similar to the one observed presently) leads to multiple streams of gas flowing towards the
black hole and creating accretion disks with angular momentum depending on the ratio of cloud and circum-nuclear disk material.
The infalling gas creates two inclined, counter-rotating sub-parsec scale accretion disks around the supermassive black hole with
the first disk forming roughly 1 Myr earlier, allowing it to fragment into stars and get dispersed before the second, counter-rotating disk forms.
Fragmentation of the second disk would lead to the two inclined, counter-rotating stellar disks which are observed at the Galactic Center.
A similar event might be happening again right now at the Milky Way Galactic Center.
Our model predicts that the collision event generates spiral-like filaments of gas, feeding the Galactic Center prior to disk formation
with a geometry and inflow pattern that is in agreement with the structure of the so called mini-spiral that has been detected in
the Galactic Center.

\end{abstract}



\section{Introduction}
\label{intro}

Recently the Milky Way Galactic Center (GC) region has again sparked a lot of interest. A number of remarkable discoveries were made, including the 
detection of two large gamma-ray bubbles \citep{2010ApJ...724.1044S}. The bubbles are perpendicular to the galactic plane, extending 50$^{\circ}$ above and below
the GC and could be linked to the GC star formation event described in this paper, as pointed out by \cite{2011MNRAS.415L..21Z}.
The model we describe in this work leads to the infall of multiple times 10$^4$ \msol\  onto the black hole within a short time. This could
be the origin of the accretion event that created the gamma-ray bubbles. The age of the  
stars observed at the GC which, according to our model formed as a result of this accretion event,
tells us that this event happened more than 6 $\pm$ 2 Myr ago (taking into account that it takes a while for
the stars to form) which coincides roughly with the age of the lobes of 10 Myr.\\

A new feature has also been detected in the GC region which previously was only identified partially and thought of as separate components:
the 100 pc elliptical and twisted ring of cold and dense molecular clouds \citep{2011ApJ...735L..33M}. 
This structure fits nicely into the 'stuff within stuff' scenario described by \cite{2010MNRAS.407.1529H} to explain the radial transport of material
through the galactic disk. According to this scenario, gravitational
instabilities generate a series of rings, bars and spirals within each others from galactic scales down to the central few parsec
region. On a smaller scale compared to the 100 pc ring we find the circum-nuclear disk (CND) which is again a ring of gas \citep{1982ApJ...258..135B}
extending roughly from 1.5 pc to around 7 pc.\\

Inside the central evacuated region of the CND the so-called mini-spiral can be found \citep{1983A&A...122..143E, 1983Natur.306..647L}, a spiral-like
structure of hot gas with a total mass of around 10$^2$ to 10$^3$ \msol\  on its way towards the black hole. At even smaller distances we find two
inclined and counter-rotating disks of young stars, extending from 0.05 pc to around
0.5 pc \citep{2009ApJ...690.1463L, 2003ApJ...594..812G, 2005ApJ...620..744G, 2006ApJ...643.1011P, 2009ApJ...697.1741B}
which are the main focus of the work presented here.\\

All the central gas is embedded into a nuclear star cluster of about 10$^7$ \msol. The cluster follows a density profile that has a core of roughly 0.5 pc
size (the stellar disks) and decreases in density up to around 30 pc \citep{2003ApJ...594..812G}.\\

Finally, very recently a small gas and dust cloud has been found, called G2. The cloud will reach its pericenter distance of
only 9 $\times$ 10$^{-4}$ pc on its orbit around the black hole as soon as in mid 2013 \citep{2012Natur.481...51G, 2013ApJ...763...78G}.
This will provide the unique opportunity for numerical simulations to make predictions of the future evolution of G2 and its interaction with the
environment that will be tested directly by observations within the next years.
One prediction that was already made is that the cloud might originate from the inner edge of one of the two stellar
disks \citep{2012ApJ...750...58B, 2012ApJ...755..155S, 2012NatCo...3E1049M}.\\

In this work we will focus on the formation of the two stellar disks, which are an important ingredient in understanding the GC region.
In \cite{2011MNRAS.412..469A} (denoted as AL1 from here on)
we tested a model that was suggested by \cite{2008ApJ...683L..37W}
to explain the formation of the progenitor accretion disk (which then should fragment into a stellar disk) by a cloud which crosses over the
black hole in parts during infall. This way parts of the cloud
pass the black hole in the clockwise direction, while the other parts move around the black hole counter-clockwise, leading to efficient redistribution
of angular momentum at the focal point of collision, downstream.
This model is able to explain the formation of a compact accretion disk.
In AL1 we used parameters for the initial cloud that are common in the GC
region \citep{2000ApJ...536..357M, 2004dimg.conf..253G}, which leads to a disk that is somewhat larger than the one observed in the GC.\\ 

The IMF of the disk-stars is quite unique, \cite{2010ApJ...708..834B}
argue that it is extremely top-heavy whereas \cite{2013arXiv1301.0540L} more recently find that it is not as
extreme as suggested by \cite{2010ApJ...708..834B} but still top-heavy compared to a standard Salpeter-slope.
This provides a hint to the origin of the stellar disks from an eccentric hot accretion disk.
A hot accretion disk would imply a larger Jeans length and thus a small number of high mass clumps
compared to a large number of small clumps for a cold disk.
The mean eccentricity of the disk stars is measured to be around 0.34 $\pm$ 0.06 \citep{2009ApJ...697.1741B}. It has already been shown that in an
eccentric accretion disk, the formation of low mass clumps can be suppressed due to tidal disruption during pericenter
passage \citep{2008ApJ...674..927A}. In addition an eccentric disk will be heated quite efficiently during pericenter passage of the gas.
These mechanisms as well as the feedback from the already formed stars which could again disrupt low mass clumps, will lead to a
strong suppression of the low mass end of the IMF and could explain the observed top-heavy IMF.\\

The papers of \cite{2008Sci...321.1060B} and very recently \cite{Mapelli:2012fb} also explain the formation of a progenitor accretion disk by
infall of a cloud with very low angular momentum into the GC. They studied the subsequent fragmentation and star 
formation inside the accretion disk and found that clouds with masses of the order 10$^5$ \msol\  can lead to a top-heavy clump mass function.\\

The one disadvantage of those models is that they can only explain the formation of a single disk. Since the stars in both disks
are almost equal in age (6 $\pm$ 2 Myr, \cite{2006ApJ...643.1011P}) they must have formed almost at the same time. The two 
stellar disks are sharing the same region, which is not possible for two gaseous counter-rotating accretion disks.
They therefore must have formed at least shortly one after another.
The basic idea here is that the first accretion disk has enough time to fragment, before the formation of the second accretion disk. The fragments and stars are
then frozen into their orbits and can share the volume of the second accretion disk, which later on also
fragments thus forming the second stellar disk.\\

The infall of a cloud with a mass larger than 10$^4$ \msol\  with a sub-parsec impact parameter is already very unlikely.
For the two disk scenario a second cloud has to fall into the central region very shortly after the first cloud, with a similar small impact parameter but inclined with respect to
the first clouds orbital plane and with opposite angular momentum.\\

In the model of \cite{2009MNRAS.394..191H} this problem is addressed by assuming the collision of two clouds further away from the black hole.
This way two streams of gas can make their way into the central region, leading to the formation of an inner compact accretion disk 
surrounded by a ring of gas.
The paper of \cite{2013arXiv1305.0012L} tries to explain the formation of two disks by the infall of a turbulent cloud.\\

In this paper we present a new model in which a single event, the interaction of a cloud with a gaseous disk around the central black hole,
naturally leads to the subsequent formation of multiple inclined and counter rotating accretion disks around the central black hole.
We present a high resolution simulation which uses initial parameters taken from a low resolution parameter study to reproduce the observed stellar disk parameters as good as possible.
This model is inspired by observational evidence which shows that a similar event to what presumably happened more than 6 Myr ago
might be happening again right now in the GC. The paper of \cite{2009ApJ...695.1477M}
shows evidence for the so called GC 20\ \kms cloud crashing into the CND, leading to the inflow of gas into the central parsec
region. This event has already been discussed before and was presented first in the paper of \cite{1993ApJ...402..173J}.
The gaseous disk in our model has similar parameters compared to the currently observed CND.\\

The extreme parameters of the two disks are quite challenging to reproduce. First they are almost equal in size, largely inclined and counter-rotating. Second the total mass in stars is quite
high with 10$^4$ \msol\  for the clockwise rotating disk and 0.5 $\times$ 10$^4$ \msol\ for the counter-clockwise disk \citep{2010RvMP...82.3121G}.\\

In section \ref{num} we present our formation model in detail, as well as a description of the algorithms we are using for our simulation and the
parts we added to the standard version of the code. We also summarize the initial parameters for the simulations.
The simulations are shown in section \ref{result}. Finally in section \ref{summary} we discuss the results and give a summary
of our work.


\section{Model and Numerical Method}
\label{num}

First we describe the basic physical ingredients of our model in section \ref{num_model}. In section \ref{num_para} we present the
hydrodynamics code we employed for running our simulations and the numerical parameters used for the simulations.
The parameters used to model the Galactic Center environment are summarized in section \ref{num_para}.


\subsection{Model}
\label{num_model}

Inspired by observations of \cite{2009ApJ...695.1477M} we simulate the collision of a cloud with a gaseous
disk (GD) around the central black hole. 
The angular momentum direction of the cloud is chosen to be opposite to the GD angular momentum. A similar model has been tested by
\cite{2001A&A...377.1016V, 2002A&A...388..128V} in their work of prograde and retrograde encounters of clouds with the observed CND to investigate
the mass infall into the central region, but they do not relate this process to the formation of the two stellar disks.\\

Since the simulated GD is destroyed in the process, the parameters do not necessarily have to be the same as the
currently observed CND. However, the general idea for the creation of the CND is that a number of clouds with similar angular momentum
are captured within the central region forming a gaseous ring within the galactic plane. Hydrodynamical simulations of this model have
been carried out by \cite{2003ANS...324..629C}.
This simple process could also have happened before, thus we assume the parameters of our GD to be similar to the CND currently
observed. Even if this is not true, it makes sense that a progenitor disk co-rotates with and lies within the galactic plane, since the overall 
source for the gas is the galactic disk.\\

In our first low resolution parameter study we have chosen a number of different inclinations between the GD orbital plane and the 
cloud orbital plane. These studies have shown that the best choice, if we want to create a small compact accretion disk, is always to 
choose those planes to be equal. This choice also makes sense given the fact that the CND (and thus also our GD)
is aligned with the galactic plane \citep{2006ApJ...643.1011P} and that the GC clouds are mostly confined to the galactic
plane \citep{2000ApJ...536..357M}.
Thus the probability for a cloud to approach the CND/GD within the same orbital plane is high.\\

As long as clouds with the same angular momentum direction as the galactic disk and thus also the GD fall in,
the GD will grow in mass.
Once in a while (a crude estimate from the two data points we have would be roughly every 10 Myr)
a cloud with opposite angular momentum direction enters the region. This leads to the destruction of the GD, resulting in
the formation of spiral-like accretion filaments that feed the Galactic Center and subsequently form
a single or multiple stellar disk(s).\\

There is a high degree of freedom in the choice of our initial parameters. The basic simplifications we assume are as follows.
The cloud is always assumed to be spherical and homogeneous. The GD is also assumed to be homogeneous with an inner cavity.
Observations show that the observed CND is a rather clumpy structure with a volume filling factor of a few percent \citep{2002A&A...384..112L}.
However, it turns out to be hard to model the clumpy structure of the CND without any additional stabilizing mechanism besides
self-gravity. The clumps are quickly tidally torn apart, reverting the disk into a homogeneous state. 
Since we do not want to deal with the question of clump stability, we use a homogeneous disk as a first approximation.
This point will be discussed further in section \ref{result_toomre} and \ref{summary}.\\

Another simplification is the assumption that the cloud and the GD always have the same mass. This is a strong restriction
and could very well be different but in order to reduce the degree of freedom we need a starting point.
In a subsequent paper we plan to study the effect of different masses and non-homogeneous density distributions.
Our current simulations do not include any feedback from the black hole or stellar feedback. Again we discuss possible implications of this simplification in section \ref{summary}.\\

In our second low resolution parameter study we always started with the GD aligned with the xy-plane, rotating counter-clockwise. For the infalling clockwise rotating cloud,
we vary the radius, mass, impact parameter and offset in z-direction. As already explained we now assume no inclination between the cloud's orbital plane and the GD orbital plane.
For the GD we vary the outer and inner radius and always set the mass equal to the cloud mass. We present the numerical values of the final set of parameters used for the
high resolution simulation in section \ref{num_gal}.


\subsection{Numerical Setup}
\label{num_para}

We carry out our hydrodynamical simulations using the N-Body Smoothed Particle Hydrodynamics (SPH) 
Code Gadget3 \citep{2001ApJ...549..681S, 2005MNRAS.364.1105S}, which makes use of the entropy-conserving 
formulation of SPH. Gadget3 improves the parallel performance of the previous version by adding a new 
domain decomposition scheme, which balances work and memory-load at the same time by a more fine grained
distribution of the particles among the CPUs, requiring slightly more communication between tasks.
For our setup, where the time-steps of particles in the vicinity of the black hole are much
smaller than in the outer parts, this already significantly improves the efficiency of the code compared
to previous versions. Additionally, we are using some modifications of the code as described in the 
following paragraphs.\\

The extremely high velocity in the vicinity of the black hole puts great stress on the domain 
decomposition. After only a few iterations, particle properties need to be communicated to other 
domains, which could reside on non-local CPUs. This strongly increased need for communication slows 
down the overall performance. Usage of the new hybrid OpenMP-MPI implementation in Gadget3
helps to overcome this problem. Here we reduce the number of MPI tasks to the number of physical 
CPUs on each node and for every MPI task we spawn additional OpenMP tasks corresponding to the number
of cores on each of the CPUs. With this approach a larger number of particles can be processed locally
without the need of MPI communication. Tests have shown that with our simulation setup we can expect a 
factor of 3-4 in performance over the standard MPI only approach.\\

Our low resolution parameter studies have shown that after crashing into the GD, the cloud is compressed
into a flat pancake like shape. This leads to strong fragmentation right
from the beginning of the simulation. In order to prevent the time step from becoming too small due to
the high density fragments we allow SPH particles to merge during the simulation. This is similar to
using sink particles although we do not replace a large number of SPH particles by a single non-SPH 
particle, we only replace two SPH particles by a single SPH particle of higher mass. Since we only allow
equal mass particles to merge and also only allow them to merge up to 3 times, the maximum mass of a high
mass SPH particle can only be 8 times higher compared to the common SPH particle mass.\\

Merging is only allowed for particles which are exactly at the minimum hydro smoothing length (see 
the end of this section). All of these particles are in a
totally collapsed region, artificially prevented from further collapse by the
choice of the minimum hydro smoothing length. The minimum hydro smoothing length sets an upper limit on
density which should always be higher than any normal region of interest inside the simulation. Only Jeans unstable
fragments which are infinitely collapsing then reach the minimum hydro smoothing length.\\

With this prescription we reduce the number of particles within a fragment. To make sure that we still 
resolve the fragment with a certain number of SPH particles, we only allow merging to start if
the number of possible candidates within the hydro smoothing length of a starting particle
(a candidate being a particle at the minimum hydro smoothing length)
is at least equal to the number of SPH neighbors which is 50 in our case.

Two particles are merged by using the center of mass position and velocity of the two particle system for
creating the replacement particle. This way linear momentum is conserved but not angular momentum. However,
due to the fact that only particles within a minimum hydro smoothing length are able to merge, the error is
extremely small and tests have shown that this does not influence the results at all. The particle merging
approach is a trade off between the sink particle formalism and a brute force calculation of the fragments.
Tests have shown that we can expect a factor of 4 or better in simulation speed compared to a brute force
calculation.\\

To account for the thermodynamics of disk formation, we use the same cooling formalism as in AL1, namely
that of \cite{2007A&A...475...37S}. It approximates cooling processes for optically thin, as well as
optically thick regions by applying an approximative radiative transfer mechanism using the diffusion
approximation. The method is suited for a wide range of temperatures (T = 10 - 10$^6$ K), a wide range of
densities ($\rho$ = 10$^{-19}$ g/cm$^3$ (and lower) - 10$^{-2}$ g/cm$^3$) and optical depths in the range
of 0 $< \tau <$ 10$^{11}$ as shown in tests performed by \cite{2007A&A...475...37S}.\\

In order to prevent small particle time steps in the vicinity of the black hole we define an accretion
radius $r_{\rm acc}$ (given in section \ref{num_gal}) within which all SPH particles are considered to be
accreted by the black hole. Those particles are removed from the simulation and will no longer take part
in the dynamical evolution. The fixed-position sink is not associated with a particle.\\

The number of SPH neighbors is set to n$_{\rm neigh} = 50 \pm 5$ and the fixed gravitational softening length
is set to $\epsilon = 10^{-3}$ pc. The hydro-smoothing length is adaptive, with the minimum allowed length
set to be equal to the gravitational softening length to prevent suppression or amplification of artificial 
fragmentation according to \cite{1997MNRAS.288.1060B}. Gadget3 treats gravitational forces as Newtonian up
to 2.8 times the gravitational softening length.\\

We assume a cloud (and thus also GD) mass of 6 $\times$ 10$^4$ \msol. The number of particles in the cloud
is 10$^6$, so that the mass of a single SPH particle is m$_{\rm SPH}$ = 0.06 \msol. Since the GD contains
the same number of particles as the cloud, the SPH particle masses are also the same. Our choice for the
particle mass leads to a minimum resolvable mass \citep{1997MNRAS.288.1060B} of 
m$_{\rm min}$ = n$_{\rm neigh} \times$ m$_{\rm SPH}$ = 3 \msol.


\subsection{Modeling the Galactic Center Environment}
\label{num_gal}

\begin{figure*}
\begin{center}
\begin{tabular}{cc}
(a) Setup & (b) Final state \\
\includegraphics[height=10cm,angle=270]{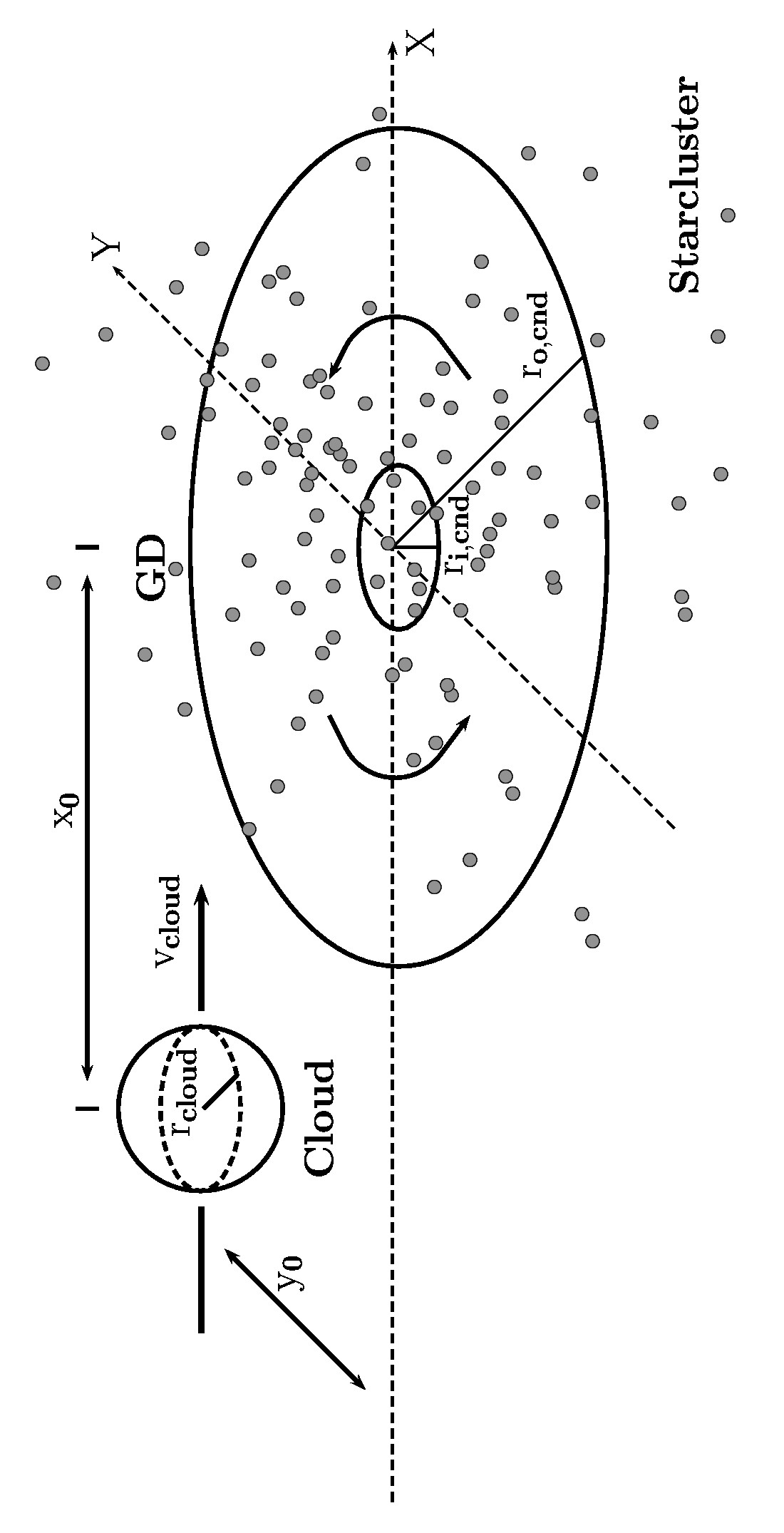} & \includegraphics[height=6cm,angle=270]{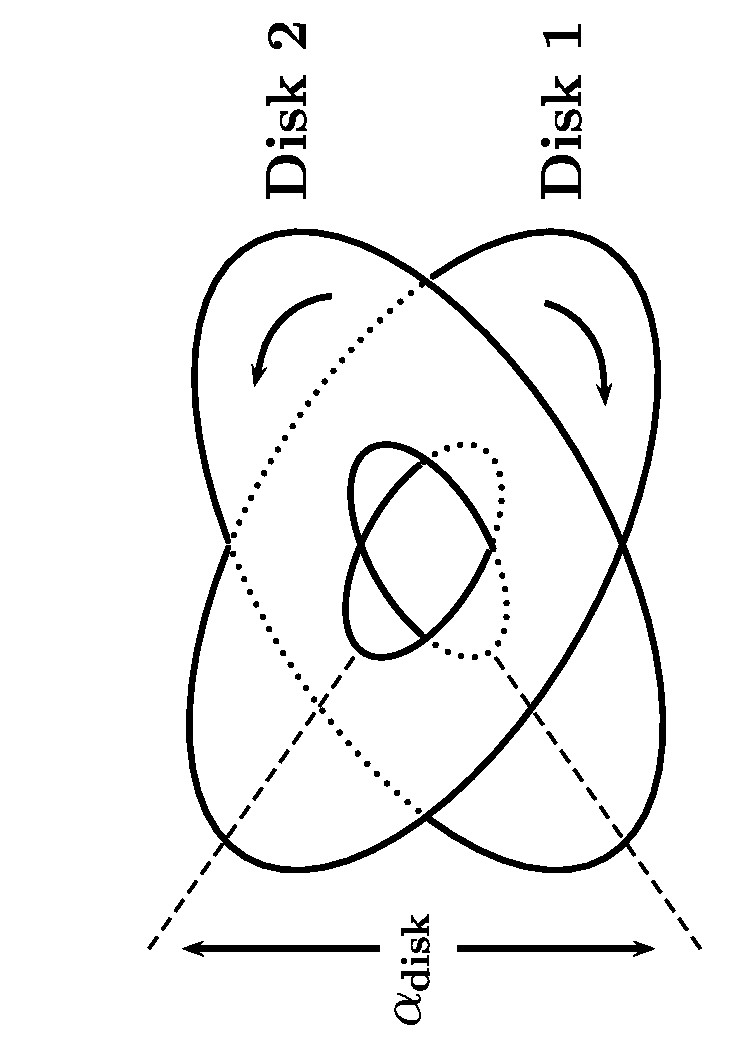}
\end{tabular}
\end{center}
\caption{
The initial setup is shown in (a) together with the simulation parameters. In (b) the two resulting accretion disks are shown
at the same time.
}
\label{setup_1}
\end{figure*}

Like in our previous work (AL1) the black hole is included as a static point mass potential, located at the origin of a Cartesian coordinate
system. However, to account for newer data from observations we increased the mass of the black hole
to 4.4 $\times$ 10$^6$ \msol\  \citep{2010RvMP...82.3121G} compared to 3.5 $\times$ 10$^6$ \msol\  in AL1.\\
For the black hole accretion radius we chose a value of $r_{\rm acc}$ = 0.1 pc.\\

In addition to the black hole potential we also include the potential of the old cluster of stars at the GC.
The density distribution is taken from observations by \cite{2003ApJ...594..812G} and has the form

$$\rho_{\star}(R) = 1.2 \times 10^6 \left( \frac{R}{0.4 pc} \right)^{-\alpha} [M_{\sun} pc^{-3}]$$

\ \\
with $\alpha$ = 2.0 for R $\geq$ 0.4 pc and $\alpha$ = 1.4 for R $<$ 0.4 pc.\\

The initial cloud has constant density and a radius of r$_{\rm cloud}$ = 2 pc.
For the mass given in section \ref{num_para} (6 $\times$ 10$^4$ \msol)
this corresponds to a density of around 10$^{-19}$ $\frac{\rm g}{\rm cm^3}$.\\

We assume an ideal gas with an adiabatic equation of state, with the cooling mechanism included as described in section \ref{num_para}.
The initial cloud and GD temperature is 100 K. The global minimum allowed temperature is also set to 100 K to account for the background UV field
from the old cluster of stars at the GC.\\

The cloud's center of mass is initially placed at an offset of x$_{\rm 0}$ = -10 pc on the x-axis and an offset
of y$_{\rm 0}$ = 5.2 pc on the y-axis.
There is no offset in z-direction, however due to the random particle distribution there is a slight (0.1 percent) abundance of mass
above the xy-plane which effectively acts like a small z-offset. 
The cloud velocity is initially set to v$_{\rm cloud}$ = 145 $\frac{\rm km}{\rm s}$ in the x-direction,
thus the cloud's orbit is clockwise around the black hole.\\

The GD is modeled as a flat disk with an inner edge of r$_{\rm i,cnd}$ = 1.8 pc and an outer edge of r$_{\rm o,cnd}$ = 7.5 pc. It is placed within the
xy-plane and rotates counter-clockwise. Fig. \ref{setup_1} depicts the simulation parameters for the initial setup and the final state.


\section{Results}
\label{result}

In this section we present the results from our high resolution simulation.
First we describe the evolution of the infall of the cloud and the interaction with the GD in \ref{result_evol}.
The amount of mass concentrated around the black hole over time is shown in \ref{result_mass}.
In \ref{result_angular} we present the angular momentum distribution of the compact accretion disks and in \ref{result_toomre}
we investigate their Toomre parameter. Finally in \ref{result_spiral} we 
shortly present results on the similarity between the inflow pattern of gas during the cloud-GD interaction and the 
observed mini-spiral.


\subsection{Evolution of the cloud-GD interaction}
\label{result_evol}

\begin{figure*}
\begin{center}
\includegraphics[width=18cm]{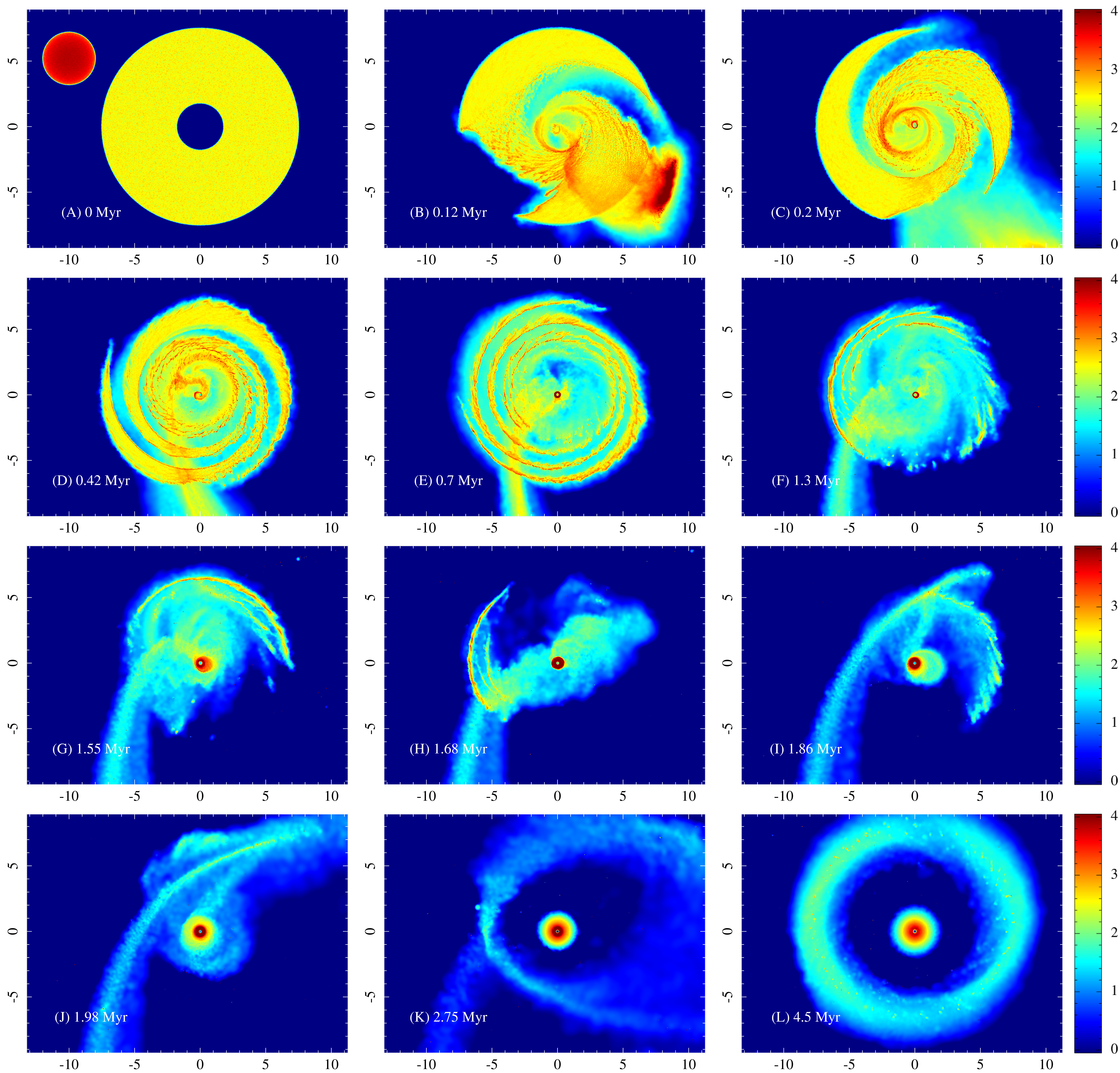}
\end{center}
\caption{
Large scale view of the cloud-GD interaction evolution. Shown is the
logarithmic surface-density in units of \sdensity \  and the length units are given in pc. 
In (a) the initial setup with the cloud approaching from the left is shown.
The cloud crashes into the GD and gets strongly compressed (b).
Next the cloud reaches the opposite side of the GD and the first low mass accretion disk forms (c).
A stream of cloud material flows backward into the inner region of the GD and starts winding up around the black hole (d).
Plot (e) shows the first high mass accretion disk at peak mass.
After the high mass accretion disk is destroyed by gas of opposite angular momentum
eating it up from the outside in, another low mass, high eccentricity and short lived accretion disk forms,
shown in (f).
A small amount of material left from the original GD starts its last orbit (g), crossing the stream
of gas from the cloud in (h). After the GD is completely destroyed, forming the second high mass accretion disk (i), the
stream of gas from the cloud can overshoot the GD region in (j). This leads to the formation of a large eccentric ring
of gas around the central accretion disk (k). This ring slowly circularizes, leading to the final steady state shown in (l).
}
\label{evol_1}
\end{figure*}

\begin{figure*}
\begin{center}
\includegraphics[width=18cm]{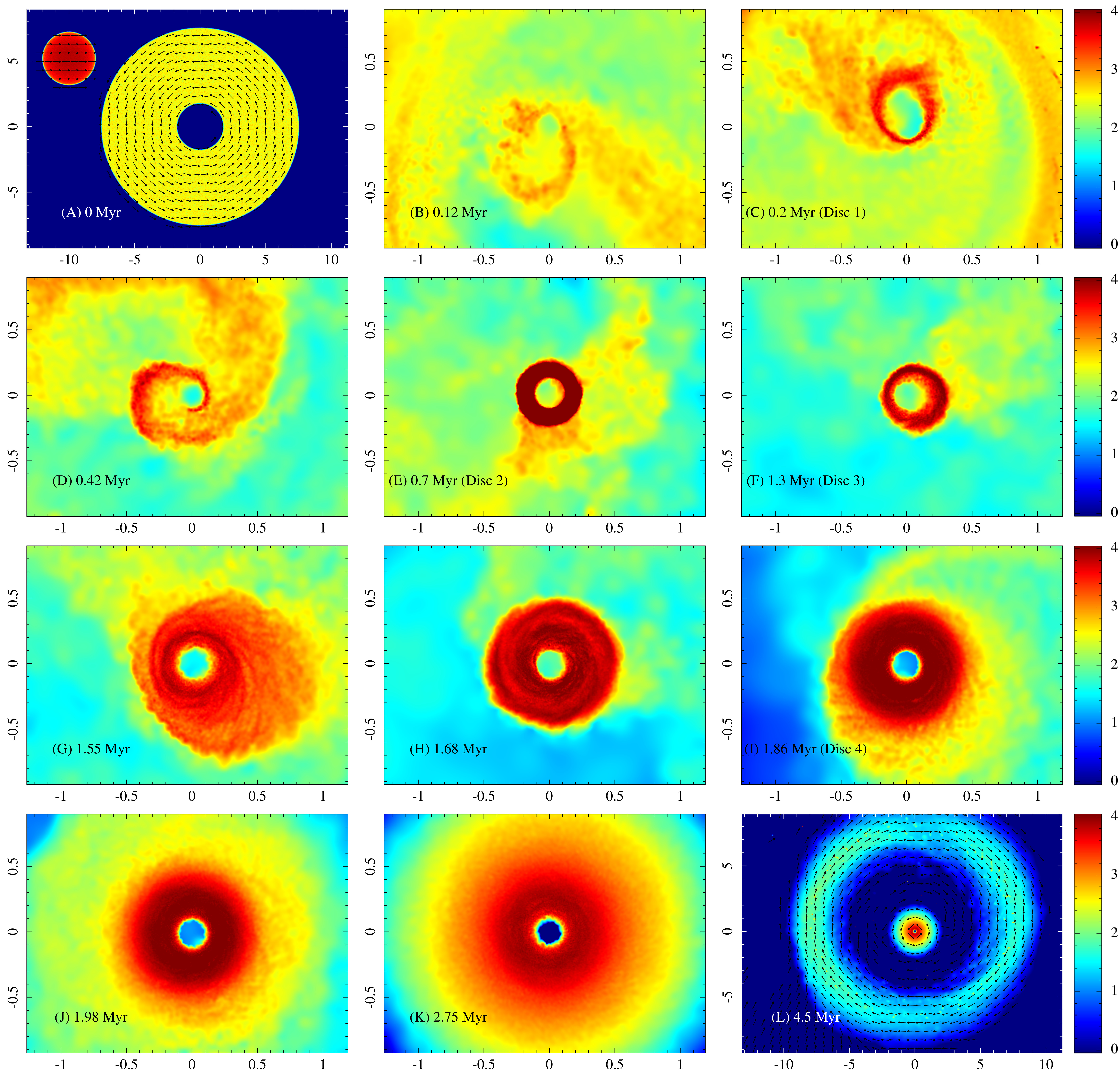}
\end{center}
\caption{
Shown is the logarithmic surface-density in units of \sdensity \  and the length units are given in pc.
In (a) the initial setup with the cloud approaching from the left is shown.
Velocity vectors are all the same length and only indicate direction.
Panels (b) to (k) give a close view of the inner region of the cloud-GD interaction evolution.
The cloud crashes into the GD and the inner cavity of the GD starts to fill up with material (b).
Formation of the first low mass accretion disk is shown in (c).
The stream of cloud material which flows backward into the inner region of the GD winds up around the black hole (d).
Plot (e) shows a close view of the first high mass accretion disk at maximum mass.
In (f) the second low mass accretion disk can be seen.
The remaining GD material is moving into the central area leading to the formation of the second high mass 
accretion disk (g, h, i).
The configuration now remains fairly steady for a long time and the disk circularizes (j, k).
At the final step (l), the outer ring of gas is almost completely circular and replaces the
original CND at lower density and with opposite angular momentum.
}
\label{evol_2}
\end{figure*}

Fig. \ref{evol_1} shows a large scale view of the evolution of the cloud-GD interaction.
The surface density is given in units of \sdensity \  and the length units are given in pc.
A closer view of the inner region can be seen in Fig. \ref{evol_2} using the same units
and time slices as Fig. \ref{evol_1}.\\

Figures \ref{evol_1} (a) and \ref{evol_2} (a) show the initial setup with the cloud approaching the counter-clockwise rotating GD from the left side.
In Fig. \ref{evol_1} (b) the cloud just crashed into the GD and reached the opposite side.
The cloud is strongly compressed and fragments are forming inside its high density region.
The formerly evacuated central region of the CND is filling up with low angular momentum
gas, better seen in Fig. \ref{evol_2} (b).\\

Following its highly eccentric orbit, the cloud moves outside the central 5 pc region (Fig. \ref{evol_1} (c)) and thus will no longer be visible on the next plots.
The cloud is tidally torn apart and a stream of gas forms that slowly falls back towards the black hole, seen on the lower right in Fig. \ref{evol_1} (c).
Fig. \ref{evol_2} (c)
shows the first sub-parsec scale accretion disk that has formed around the black hole at its peak mass.
The disk has very low mass (6.5 $\times$ 10$^2$ \msol), is short lived (roughly 0.1 Myr) and highly eccentric. 
There is still a lot of material inside the GD, which is now spiraling towards the center.\\

The tight stream of gas entering from the bottom of Fig. \ref{evol_1} (d) still originates from the cloud.
This stream mixes with the ambient gas from the GD (Fig. \ref{evol_2} (d)), by this loosing angular momentum, and winds up around the black hole.
This leads to the formation of the second accretion disk shown in Fig. \ref{evol_1} (e) and
Fig. \ref{evol_2} (e).\\

The disk has a high mass (1.35 $\times$ 10$^4$ \msol) and exists for a long time (roughly 0.5 Myr).
Besides forming the first high mass accretion disk the stream also collides
with material inside the GD, thus slowly reducing the angular momentum (due to the collision with material of opposite angular momentum) of the GD material, moving it further towards the center.\\

The high mass accretion disk is now slowly eroded by GD gas with opposite angular momentum falling in at the outer edges.
This destroys the disk from the outside in.
A detailed study of the evolution of nested accretion disks with
opposite angular momentum close to supermassive black holes has been carried out by \cite{2012MNRAS.422.2547N}).
They find that accretion rates can increase more than 100 times above the accretion rate of a single planar 
viscously evolving accretion disk. This provides a very efficient way to feed SMBHs.\\

After the disk has vanished another short lived (roughly 0.1 Myr), low mass (9.5 $\times$ 10$^2$ \msol) and high eccentricity accretion
disk forms (Fig. \ref{evol_1} (f) and Fig. \ref{evol_2} (f)). The GD is now almost completely destroyed by the stream of gas flowing back in from the cloud.\\

Transition into the final steady state is shown in the series of plots (g-l) in Fig. \ref{evol_1} and Fig. \ref{evol_2}.
In Fig. \ref{evol_1} (g) the last material remaining in the GD starts its last 
orbit, crossing the stream of backwards flowing gas from the cloud in Fig. \ref{evol_1} (h).
All this material now winds up around the central black hole and 
forms the second high mass accretion disk (5.65 $\times$ 10$^3$ \msol), presented in Fig. \ref{evol_1} (i) and Fig. \ref{evol_2} (i) at peak mass.\\

When most of the GD gas has fallen inwards the gas from the clockwise rotating stream, resulting from the disrupted gas cloud cannot lose angular momentum
anymore and feeding of the central accretion disk stops. Fig. \ref{evol_1} (j) shows the stream of gas
from the cloud overshooting the GD region since there is no more material from the GD left to collide with.

Finally in Fig. \ref{evol_1} (k) the stream of gas from the cloud forms a large, eccentric ring of gas around the inner compact accretion
disk. The ring slowly dissipates kinetic energy and circularizes, leading to the final steady state at 4.5 Myr shown in Fig. \ref{evol_1} (l).
An interesting feature of this ring is that it almost replaces the original GD (concerning location and width, although with a much lower density)
but now rotates clockwise (compare velocity vectors in Fig. \ref{evol_2} (a) and (l)), as it is comprised almost completely of cloud material.
The subsequent infall of additional clouds with the same angular momentum as the initial cloud could now add to the mass of
this new GD eventually building up the currently observed CND.
The infall of a cloud with opposite angular momentum could trigger the next accretion phase
onto the supermassive black hole, which we might be witnessing currently \citep{2009ApJ...695.1477M}.
The inner accretion disk is no longer affected by the infall of gas and remains more or less steady (Fig. \ref{evol_2} (j, k)).

\begin{figure*}
\begin{center}
\begin{tabular}{cc}
(a) 0.7 Myr & (b) 1.86 Myr\\
\includegraphics[width=8.5cm]{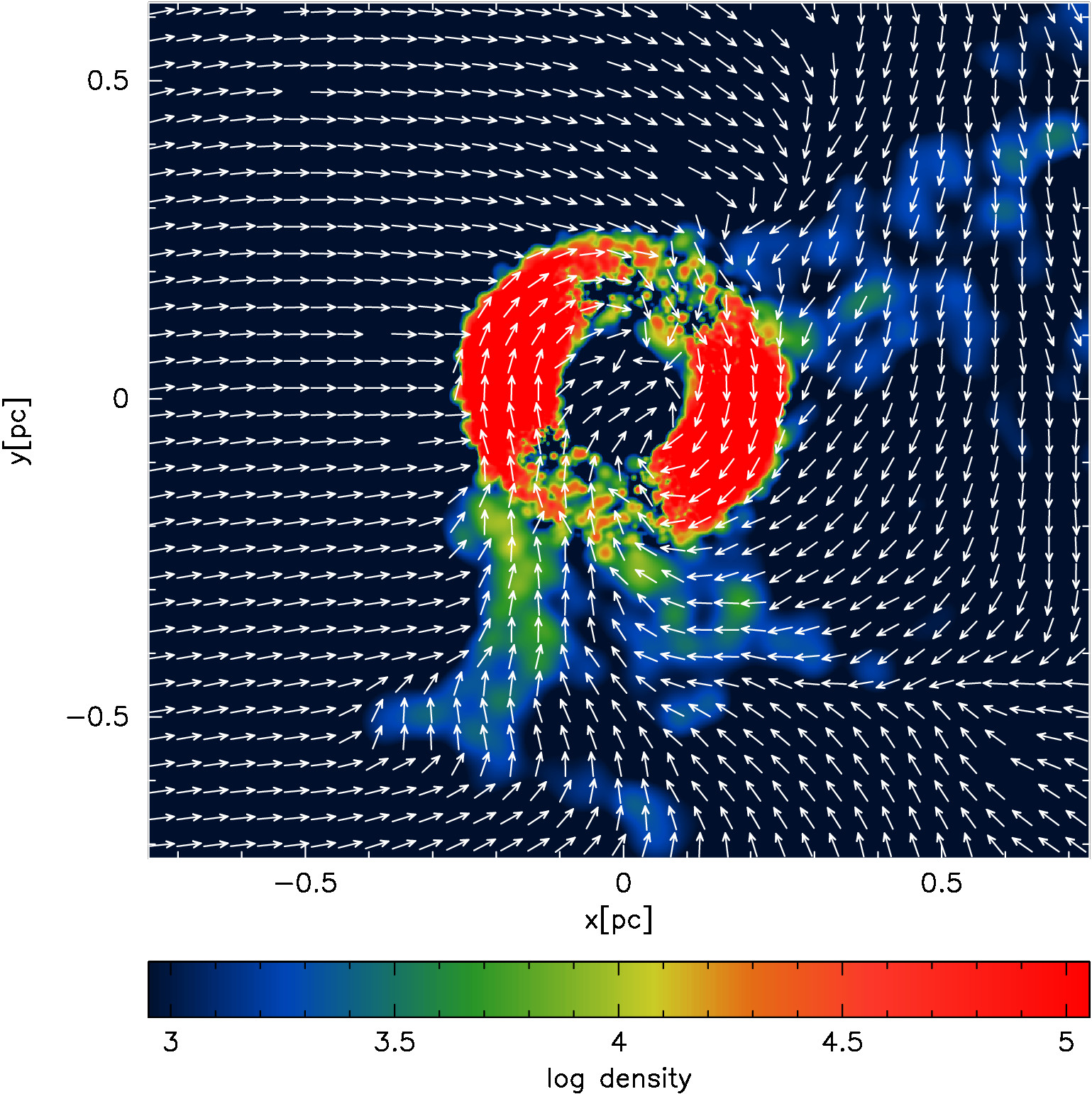} & \includegraphics[width=8.5cm]{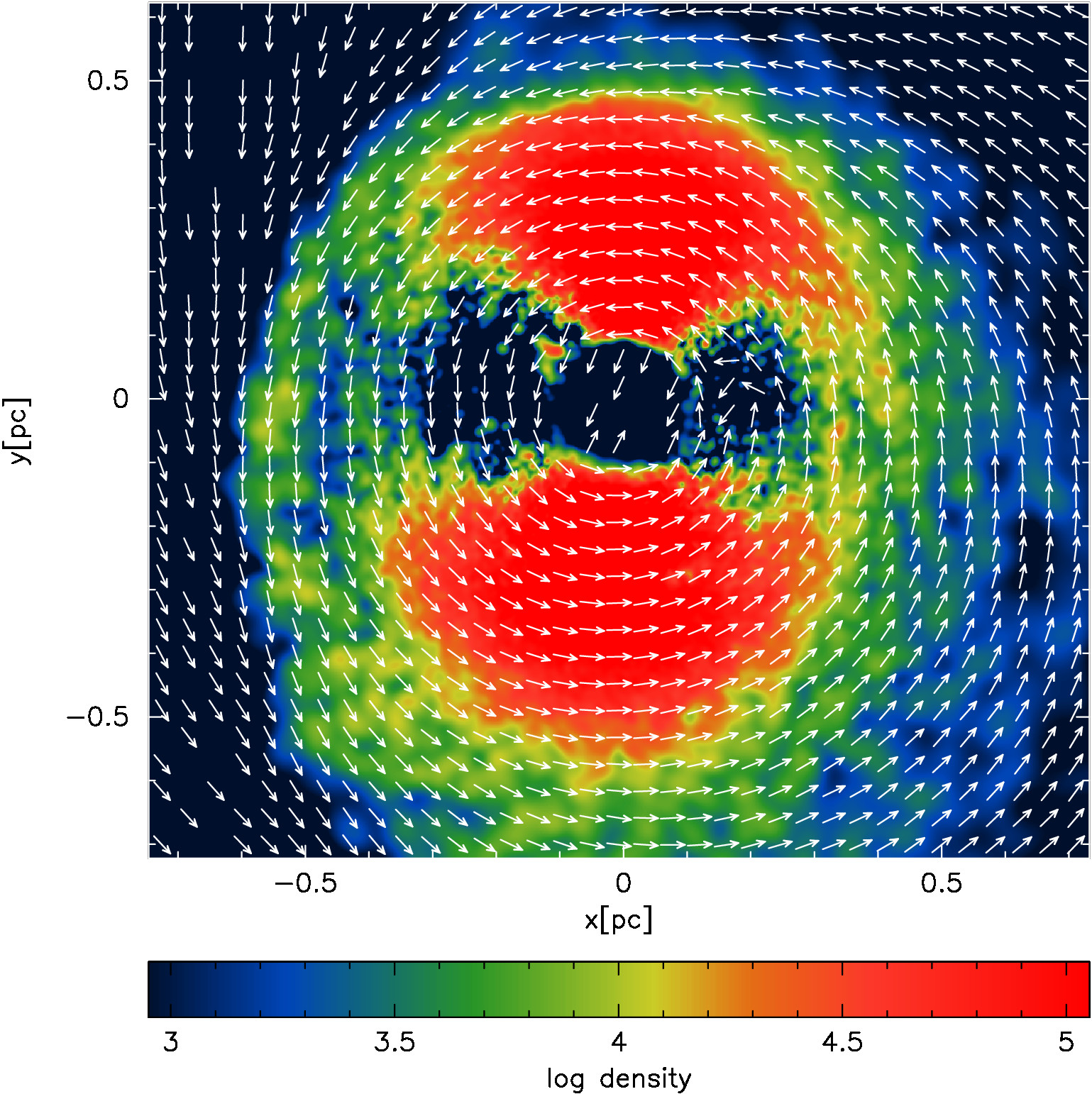}
\end{tabular}
\end{center}
\caption{Density cut in the z=0 pc plane in units of \density.
We compare the density in the z=0 plane of the two high mass accretion disks forming at 0.7 Myr (a) and 1.86 Myr (b).
Velocity vectors are all the same length and only indicate direction.
The disks are counter-rotating, inclined with respect to the xy-plane and with respect to each other.
}
\label{evol_3}
\end{figure*}

A comparison of the density of the two high mass accretion disks in the z=0 pc plane is shown in Fig. \ref{evol_3}.
Several points which we investigate in more detail later are already visible from these plots.
First the disks are clearly counter-rotating, the first accretion disk rotates clockwise, the second accretion disk counter-clockwise.
Second the disks are inclined with respect to the xy-plane. They are rotated along the connection of the high density parts
in Fig. \ref{evol_3}, almost horizontal for (a) and almost vertical for (b).
The disks can not be inclined in the same way, since the axes of rotation are nearly perpendicular to each other.


\subsection{Amount of mass within the central region}
\label{result_mass}

\begin{figure}
\begin{center}
\includegraphics[width=8.5cm]{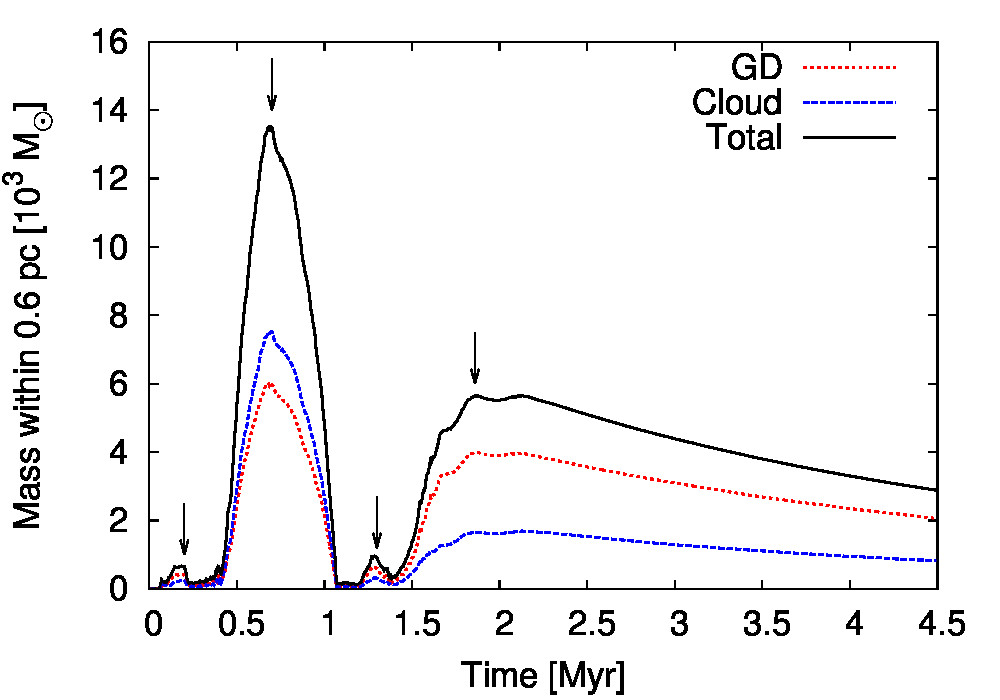}
\end{center}
\caption{Amount of mass within the central 0.6 pc versus simulation time. The black (solid) line shows the total mass, the blue (dashed) line
the contribution from the cloud and the red (dotted) line the contribution from the GD to the total mass. The arrows indicate the points
at which we plotted the disks in Fig. \ref{evol_2}.}
\label{mass_1}
\end{figure}

\begin{figure}
\begin{center}
\includegraphics[width=8.5cm]{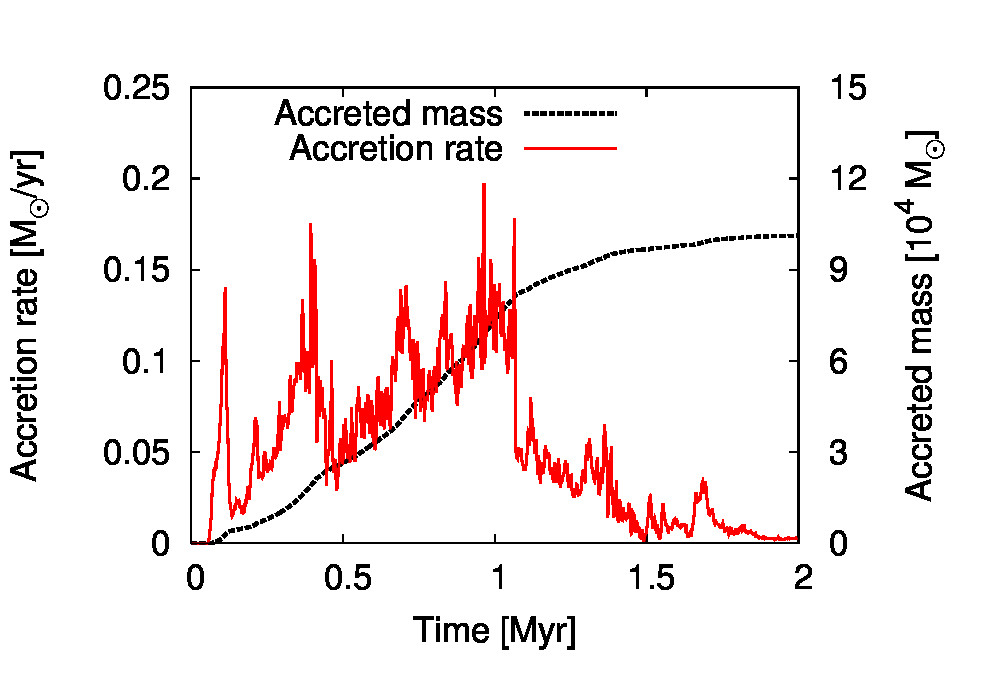}
\end{center}
\caption{
Accretion rate (red, solid line) and total accreted mass (black, dashed line) over time.
The plot is cut beyond 2 Myr since the accretion rate stays near zero until the end of the simulation at 4.5 Myr.
}
\label{mass_2}
\end{figure}

We now look at the amount of mass within the central 0.6 pc around the black hole against time. This radius is taken simply because it is a bit further out than the radius of the largest accretion disk that formed.
We distinguish between contributions from the cloud and the GD.
This provides a good first hint for the formation of the various disks around the black hole and about their sense of rotation. A peak in the mass distribution represents a compact accretion disk.
If the contribution to the mass of the disk is dominated by the cloud we expect the disk to be rotating clockwise, if the GD dominates we
expect the disk to be rotating counter-clockwise. In our low resolution parameter study this provided a quick way to identify good models.

The arrows in Fig. \ref{mass_1} indicate the points at which we plotted the disks in section \ref{result_evol}.
First we can see that the two low mass disks at 0.2 Myr and 1.3 Myr are indeed very short lived and contain almost no mass
above the ambient level, thus they are completely negligible.
On the other hand the first high mass disk contains around 11 percent (1.35 $\times$ 10$^4$ \msol) and the second high mass disk around 4 percent
(5.65 $\times$ 10$^3$ \msol) of the total initially available mass (1.2 $\times$ 10$^5$ \msol).
The lifetime of the first disk of around 0.5 Myr should be long enough for the disk
to fragment if it was unstable. A typical value for the fragmentation time of such a disk is 0.2 to 0.3 Myr as found in the
simulations of \cite{2008Sci...321.1060B}, \cite{Mapelli:2012fb} and our own accretion disk fragmentation studies.\\

The main contribution to the mass of the first disk comes from the cloud, thus the disk is rotating clockwise. Although the disk is smaller
compared to the second high mass disk, it contains more mass and thus is closer to being able to fragment as we will show in \ref{result_toomre}.
For the second disk, the contribution to the mass is higher from the GD leading to a counter-clockwise rotating disk.\\

The mass in our disks is opposite to what is seen in observations. The observed larger disk has double the mass (10$^4$ \msol) in stars compared
to the smaller disk (0.5 $\times$ 10$^4$ \msol). In our simulation the smaller disk contains roughly three times the mass of the larger disk
(Note that we compare the observed stellar disks to the simulated gaseous accretion disks).
However, we regard this as a matter of fine tuning, like a more fine grained search over the initial cloud radius or GD size,
which is beyond this first approach.\\

Fig. \ref{mass_2} shows the accretion rate and total accreted mass over time.
In total around 10$^5$ \msol\ are accreted over 2 Myr and the mean accretion rate is 
roughly 0.05 \msol/yr. Accretion in our case represents the flow
of material below the accretion radius $r_{\rm acc}$ and is not connected to a black hole accretion rate.
The material could settle into an actual black hole accretion disk at much smaller radii. The Lense-Thirring effect 
aligns the disk angular momentum with the black hole angular momentum, thus removing the accretion rate amplifying effect
of counter-rotating gas falling in at later times. The accretion disk then could evolve viscously, thus feeding the black hole
at lower rates compared to Fig. \ref{mass_2}.


\subsection{Angular momentum distribution and disk orientation}
\label{result_angular}

\begin{figure*}
\begin{center}
\begin{tabular}{cc}
(a) & (b)\\
\includegraphics[width=8.5cm]{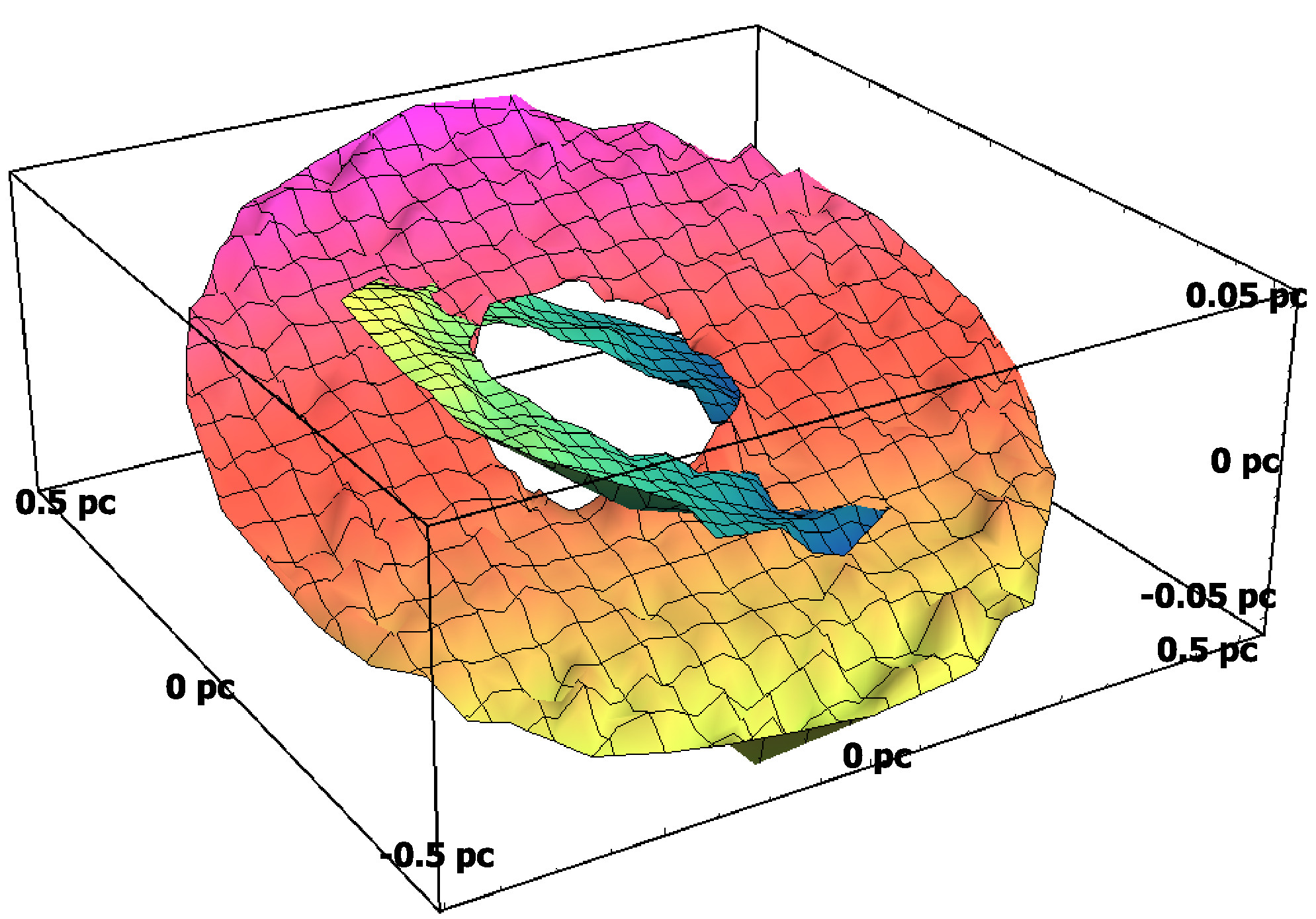} & \includegraphics[width=8.5cm]{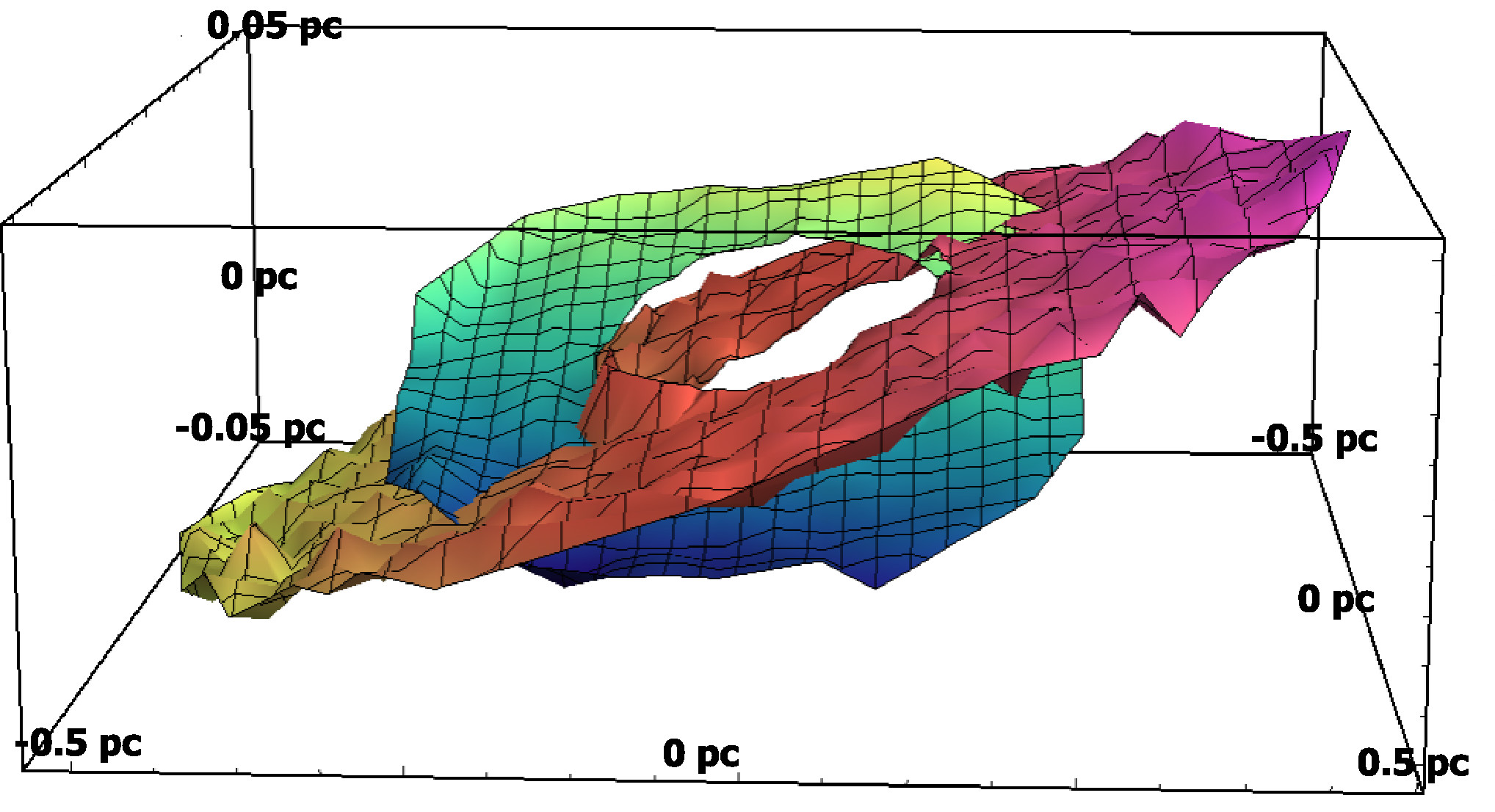}
\end{tabular}
\end{center}
\caption{
The two high mass accretion disks combined in a single plot, fitted at midplane at the times shown in Fig. \ref{evol_2} (e) at 0.7 Myr and Fig. \ref{evol_2} (i) at 1.86 Myr
from different viewing angles.
}
\label{angular_1}
\end{figure*}

\begin{figure*}
\begin{center}
\begin{tabular}{cc}
(a) 0.7 Myr & (b) 1.86 Myr\\
\includegraphics[width=8.5cm]{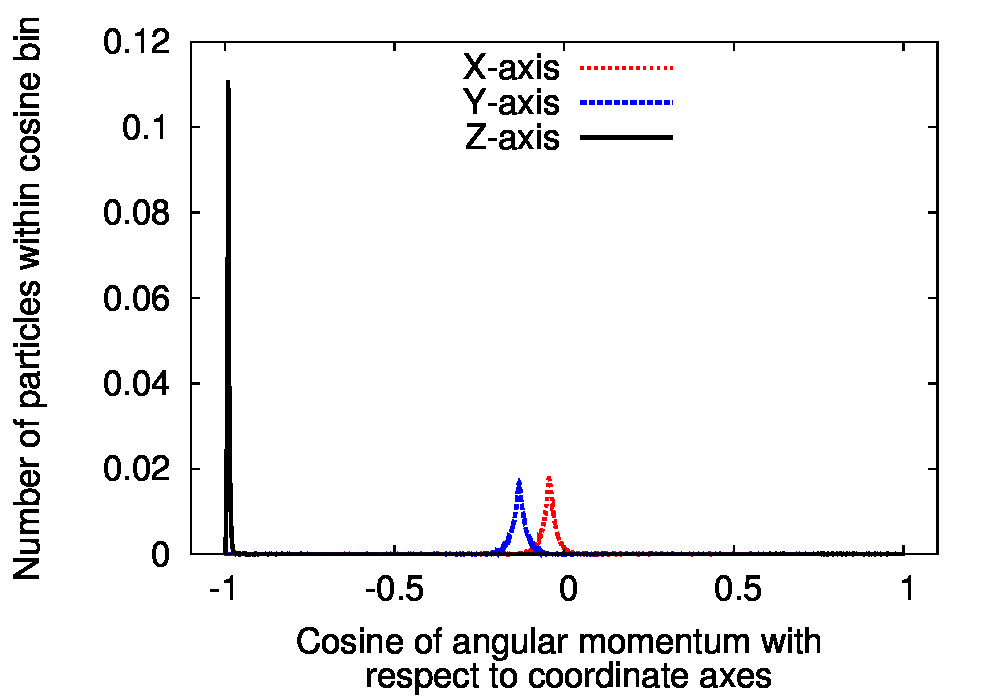} & \includegraphics[width=8.5cm]{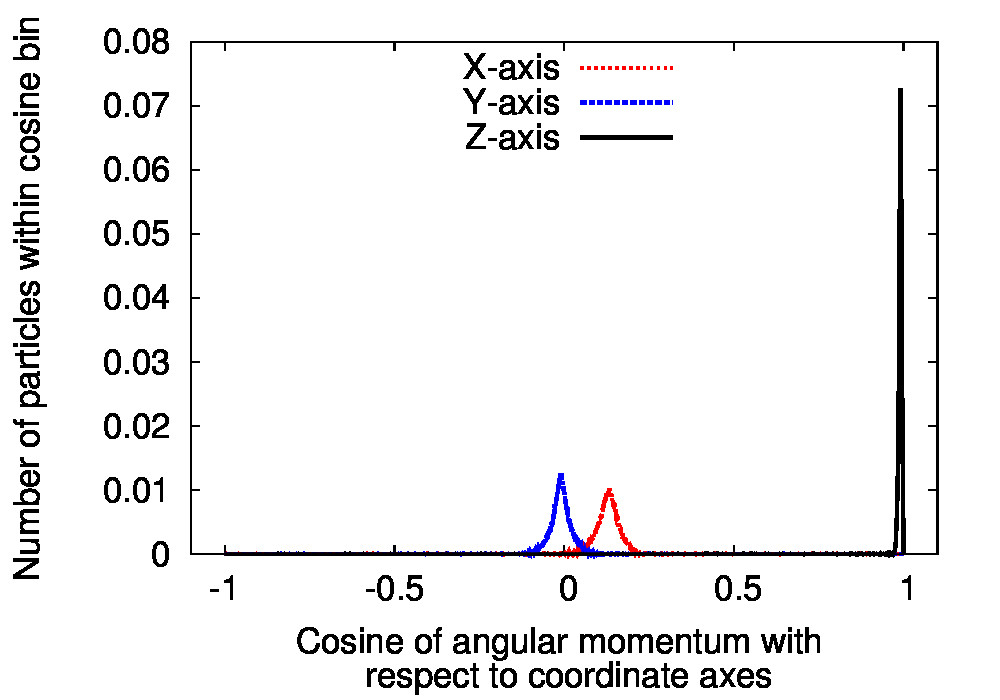}
\end{tabular}
\end{center}
\caption{Histogram of the cosine of angular momentum with respect to the x,y and z-axis. 
The black (solid) line shows the cosine of angular momentum with respect to the z-axis, the blue (dashed) line with respect to the y-axis
and the red (dotted) line with respect to the x-axis.
The two peaks near -1  and 1 for the first (a) and second (b) high mass accretion disks show that the z-component of angular
momentum is opposite for the two disks.
}
\label{angular_2}
\end{figure*}

\begin{figure*}
\begin{center}
\begin{tabular}{cc}
(a) & (b)\\
\includegraphics[width=8.5cm]{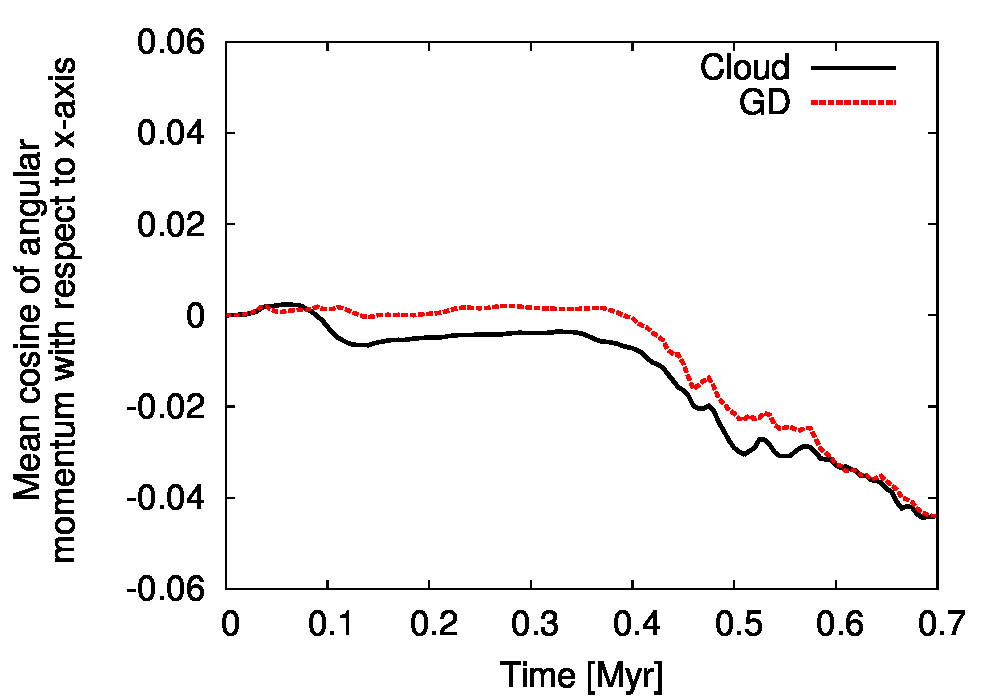} & \includegraphics[width=8.5cm]{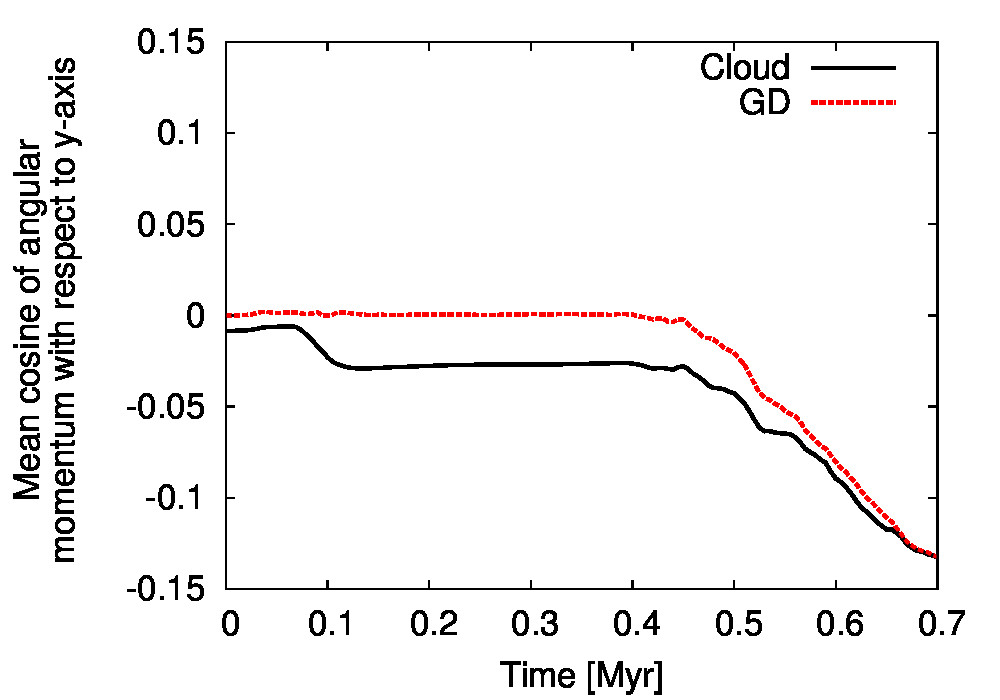}\\
(c) & (d)\\
\includegraphics[width=8.5cm]{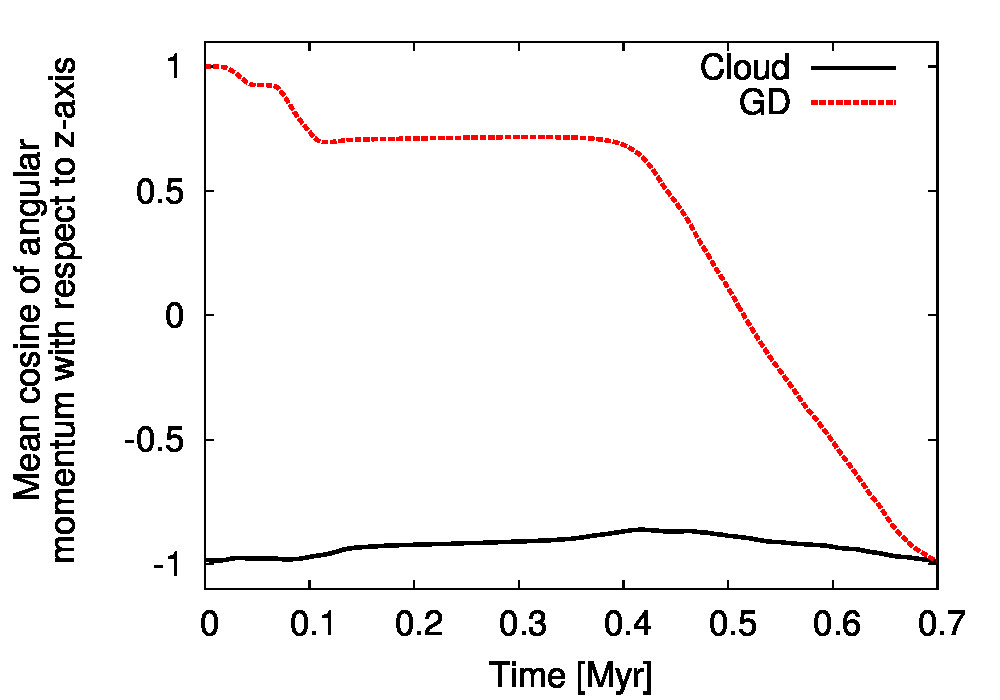} & \includegraphics[width=8.5cm]{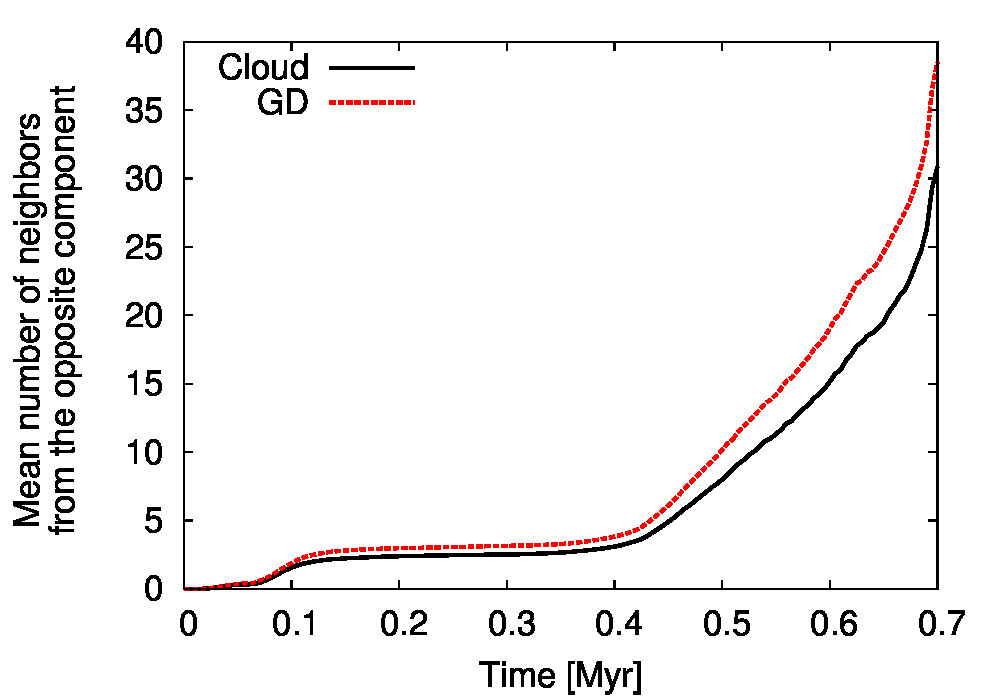}
\end{tabular}
\end{center}
\caption{Shown in (a), (b) and (c) is the
mean cosine of angular momentum with respect to x,y and z-axis of the subset of particles from the first high mass accretion disk
over time. The black (solid) line shows the angular momentum orientation for the cloud particles and the red (dashed) line for the GD particles.
In (d) we show the mean number of neighbors within the hydro-smoothing
length from the opposite component for each of the components (GD and cloud). Again only the subset of particles inside the first high mass
accretion disk at 0.7 Myr is shown. This provides a way to see how well ``mixed'' the two components are.
The mean number of neighbors from the GD for the cloud material is shown as black (solid) line and the mean number of neighbors from the cloud
for the GD material is shown as red (dashed). There is a clear correlation between mixing and angular momentum distribution.
}
\label{angular_3}
\end{figure*}

To visualize the disk configuration better, we fitted each of the high mass accretion disks with a thin sheet
at the times shown in Fig. \ref{evol_2} (e) and Fig. \ref{evol_2} (i), positioned at the
midplane of the disks. Then we combined the results into a single plot, shown in Fig. \ref{angular_1} (a) and (b) for different viewing angles.
Clearly the disks are inclined with respect to each other and share the same volume. 
The smallest angle between the disks (see Fig. \ref{setup_1} (b)) is roughly $\alpha_{\rm disk}$ = 15$^\circ$ in our case, compared to 70$^\circ$ inferred from observations \citep{2009ApJ...697.1741B}.
For comparison, the best fit of the observational data can be seen in Fig. 20 in \cite{2009ApJ...697.1741B}.
Although the second high mass disk is weakly warped in our case, the warp is not as strong as the observed one.
However, there are also cases in the low resolution runs in which a highly warped disk was produced.\\

In Fig. \ref{angular_2} we show a histogram of the binned cosine of the angular momentum vector with respect to the x,y and z-axis for all
the gas particles within the central 0.6 pc from the black hole.
The first disk, shown in Fig. \ref{angular_2} (a) has a strong peak in the z-component near -1, thus the z-component
of angular momentum is anti-parallel to the z-axis for all the gas inside the disk.
On the other hand the second high mass disk peaks near 1, thus the z-component of angular momentum is parallel to the z-axis.\\

On the plane of sky, the GD is seen almost edge on from earth. One of the observed stellar disks is seen almost completely 
face on in the plane of sky. The second stellar disk is again seen almost edge on, but perpendicular to the GD. 
This configuration is not too far from what our simulation shows. None of our simulated disks lies within the plane of the GD, both of
them are inclined with respect to the GD and with respect to each other, making our model an attractive formation mechanism for the stellar disks.\\

To show that the disk angular momentum distribution is determined by the mixture of cloud and GD material we now look only at the first
high mass accretion disk.
We isolate all the particles within the disk at 0.7 Myr and trace them back to the initial setup. For every snapshot 
we then plot the mean of the cosine of angular momentum with respect to the x,y and z-axis of those isolated particles.\\

To measure how strongly the material is mixed, we determine for every snapshot the mean number of particles from the GD 
within the hydro-smoothing length of every cloud particle, as well as the other way around.
If a particle is within the hydro-smoothing length of another particle it contributes to its density, which by definition means
that the two particles are interacting (or more sloppy formulated ``mixed'').\\

The distribution of the mean cosine of angular momentum over time obtained this way is shown in Fig. \ref{angular_3} (a), (b) and (c).
For comparison the mean number of neighbors from the opposite component can be seen in Fig. \ref{angular_3} (d).
In the beginning, cloud and GD are well separated, with no particle from the opposite component nearby and with the angular momentum
only aligned parallel and anti-parallel to the z-axis. At around 0.12 Myr the cloud reaches the opposite side after crashing into the GD.
The interaction of cloud and GD leads to mixing of the particles and now each of the components has a small number of neighbors from the opposite 
component.
This is also reflected in the mean cosine distribution, taking the z-component (Fig. \ref{angular_3} (c)), one can clearly see
that a fraction of the GD material is affected and now co-rotates with the cloud material.

At around 0.4 Myr formation of the accretion disk starts. Now the number of neighbors from the opposite component steeply rises for 
both the GD and the cloud. This is mirrored in the angular momentum distribution, again best seen in the z-component (Fig. \ref{angular_3} (c)).
The plots show that there is a strong correlation between mixing of the two components and the angular momentum distribution.
The contribution of GD material to the cloud material reduces its angular momentum, so that it can settle into a tight orbit around
the black hole. 
Fig. \ref{angular_3} (b) shows that the particles from the cloud which end up inside the accretion
disk already have a small y-component
in their angular momentum, which leads to the inclination of the accretion
disk with respect to the xy-plane.
This asymmetric distribution originates from the small imbalance of mass above and below the xy-plane due to the random
particle setup, which effectively acts like a small cloud offset in z-direction. A higher mass imbalance or a non zero cloud offset in z-direction
would increase the inclination, which we confirmed in test-simulations.


\subsection{Toomre parameter and disk fragmentation}
\label{result_toomre}

\begin{figure}
\begin{center}
\includegraphics[width=8.5cm]{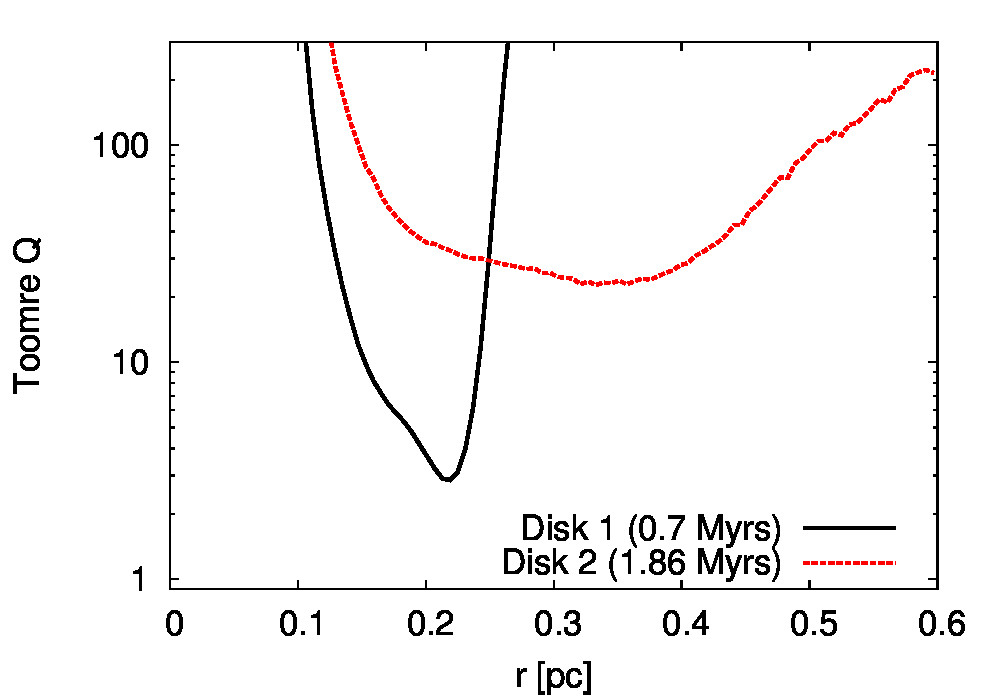}
\end{center}
\caption{Toomre Q at peak mass for both compact accretion disks. Plotted is the Toomre parameter in radial bins against
distance from the black hole. The black (solid) line shows the first high mass accretion disk which formed around 0.7 Myr,
the red (dashed) line the second high mass accretion disk which formed around 1.86 Myr.}
\label{toomre_1}
\end{figure}

\begin{figure*}
\begin{center}
\begin{tabular}{cc}
(a) Overview at 1.3 Myr & (b) Accretion disk at 1.3 Myr \\
\includegraphics[width=8.5cm]{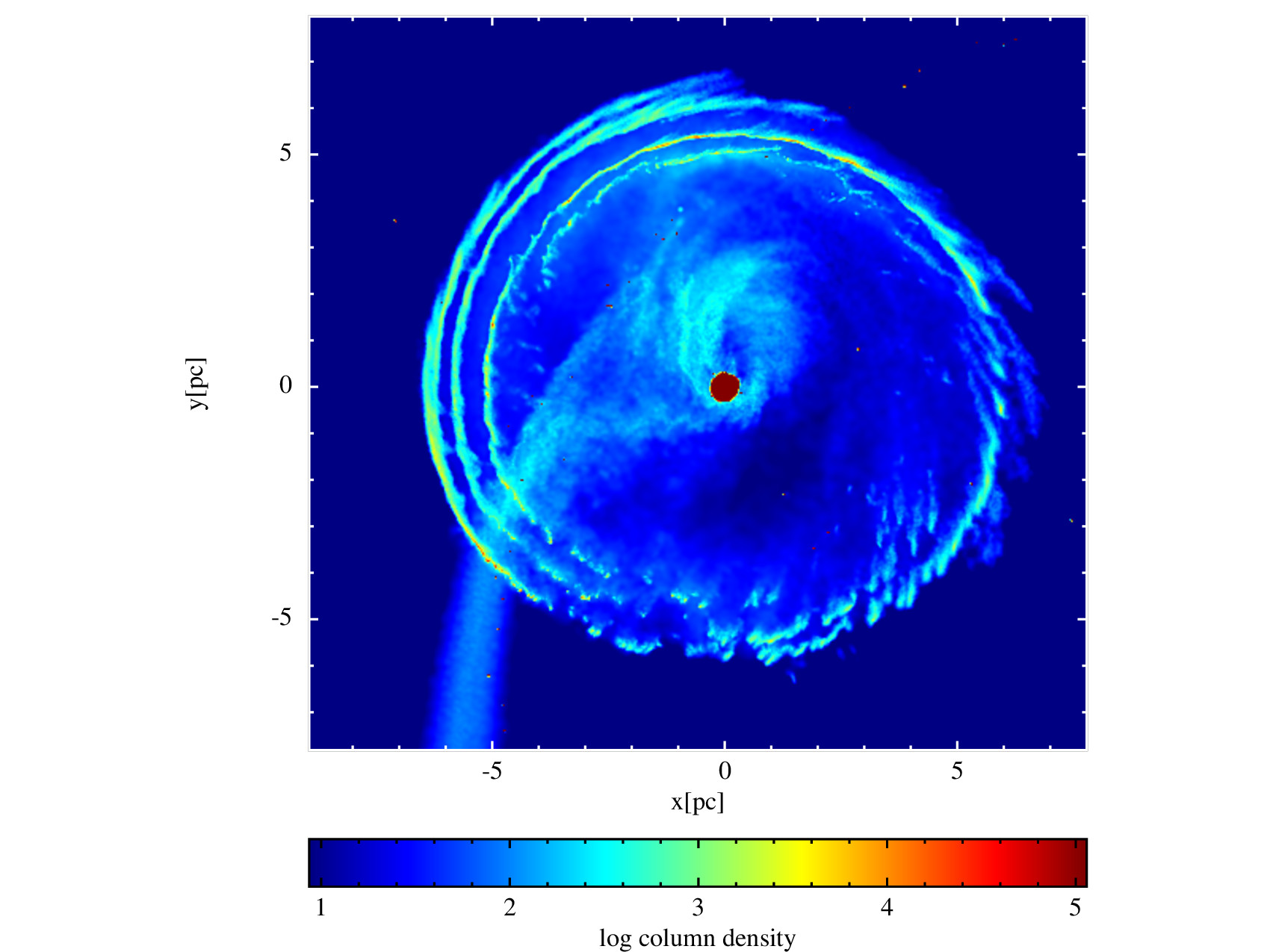} & \includegraphics[width=8.5cm]{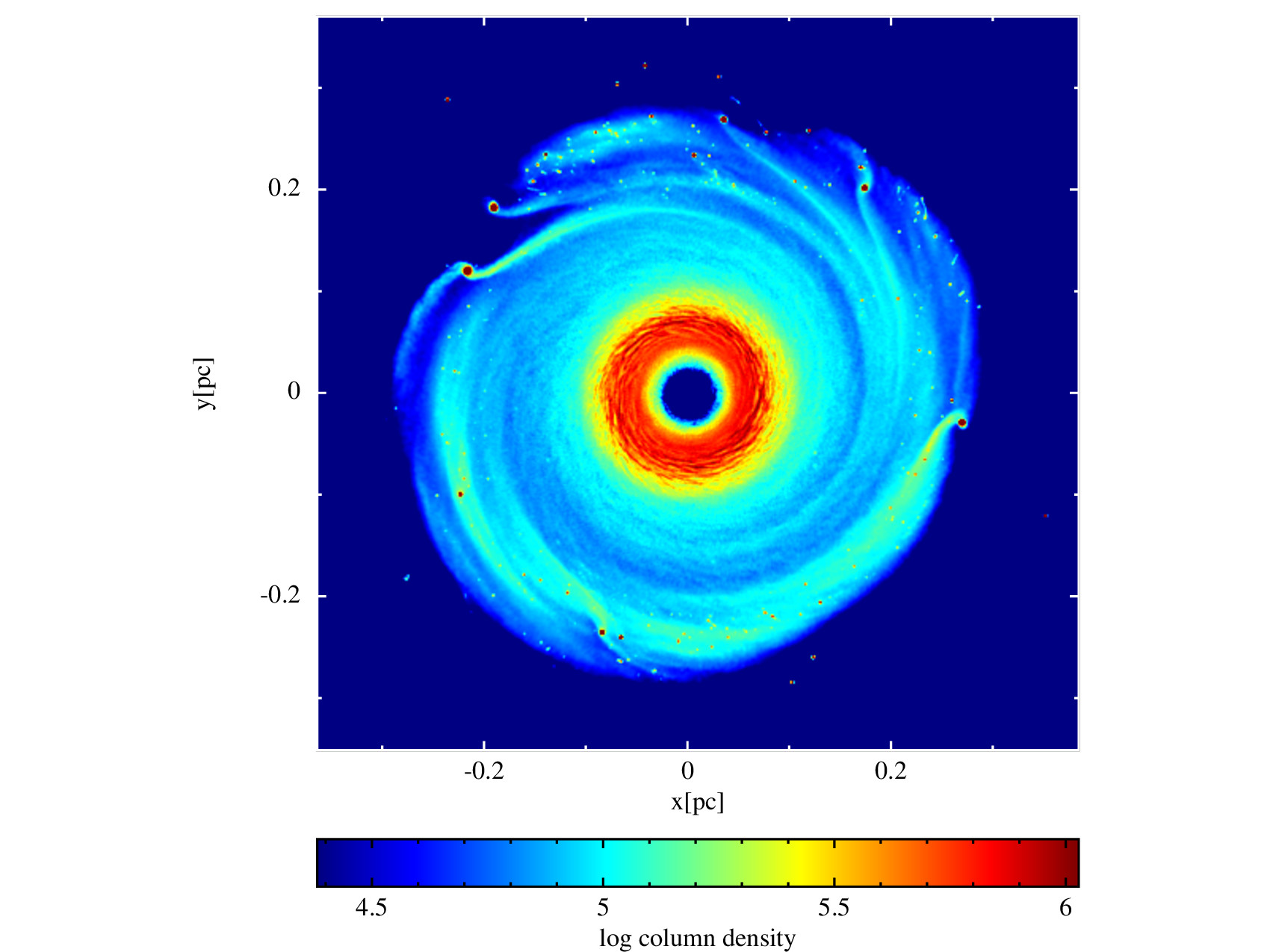}
\end{tabular}
\end{center}
\caption{Logarithmic surface-density in units of \sdensity.
A test case with different initial conditions from the simulation presented throughout the paper 
leads to a high mass accretion disk that breaks up into fragments.
}
\label{toomre_2}
\end{figure*}

After we have formed gaseous accretion disks, the next
important step is disk instability. Our scenario assumes that the first high mass accretion disk fragments into cores before the gaseous
component gets destroyed and the second high mass accretion disk forms, leaving a collisionless disk of star forming cores from the first accretion
disk. After that the second accretion disk fragments, leaving two inclined disks of cores which turn into stars and
probably undergo further evolution (e.g. change of inclination, warping) leading eventually to the stellar disks observed today.\\

To investigate the fragmentation behavior we look at the Toomre instability parameter for our two high mass accretion disks, shown in Fig. \ref{toomre_1}.
For this we rotate each of the disks into the xy-plane and calculate Q for different radial bins. 
The Toomre parameter for the first high mass accretion disk, plotted in black (solid), drops to Q=2.8 at around
0.22 pc distance from the black hole, which is still in the stable regime.
An even higher value is seen for the second high mass accretion disk shown in red (dashed) in Fig. \ref{toomre_1}. Here Q=23 represents the lowest 
value reached at around 0.36 pc. Thus both our disks are stable against fragmentation.\\

This is not unexpected since we set up our initial conditions with a rather low total mass. As already pointed out in the introduction, simulations
which use a cloud of low angular momentum captured by the black hole need up to 10$^5$ \msol\  of mass in the cloud to explain the top-heavy IMF in a single disk.
In our case we have a starting mass of 1.2 $\times$ 10$^5$ \msol\  that is potentially available for a total of two disks.\\

The low initial mass comes from considerations about cloud and GD stability. We already discussed the problem of stabilizing clumps
within the GD/CND in section \ref{num_model}. At high masses corresponding to several times 10$^5$ \msol\  the disk and the GD quickly become
unstable and start fragmenting.\\

Observations indicate that there are around 10$^4$ to 10$^5$ \msol\  of gas inside the currently observed CND and that the CND is a transient
structure (t$_{\rm life} < 10^5$ yrs) \citep{1987ApJ...318..124G}. However there are also models which assume the CND to be a rather stable structure
(t$_{\rm life} > 10^7$ yrs) \citep{2003ANS...324..613V}. Our simulations start with a progenitor of the current CND, so the mass does not necessarily have
to be the same as the current CND mass, leaving us some freedom in the choice. 
The issue of how to stabilize the GD or a cloud goes beyond the scope of our work.
Hence, our choice of parameters is a trade-off between a GD and cloud stable enough to survive the first encounter 
given our limited physics and a preferably high initial mass.
Turbulent, rotating clumps inside the GD together with a magnetic field could be ingredients needed to stabilize such a system.\\

However, even at our current initial mass we
were able to find initial conditions which lead to enough mass inside the first high mass accretion disk so
that it becomes unstable.
Fig. \ref{toomre_2} shows such a simulation in which the first high mass accretion disk that 
forms quickly becomes unstable and breaks up into cores. Unfortunately this simulation evolves very slowly 
and we are not yet far beyond the point at which the simulation is shown.
The results of this improved model will be shown in a subsequent paper.\\

A number of clumps, originating from the cloud after crashing into the GD and fragmenting, are spread out over a large area (up to 40 pc) around
the central black hole. Those clumps might represent the population of young massive stars at radii larger than 0.5 pc that were recently found
by \cite{2013A&A...549A..57N}.


\subsection{Formation of the mini-spiral}
\label{result_spiral}

\begin{figure}
\begin{center}
\includegraphics[width=8.5cm]{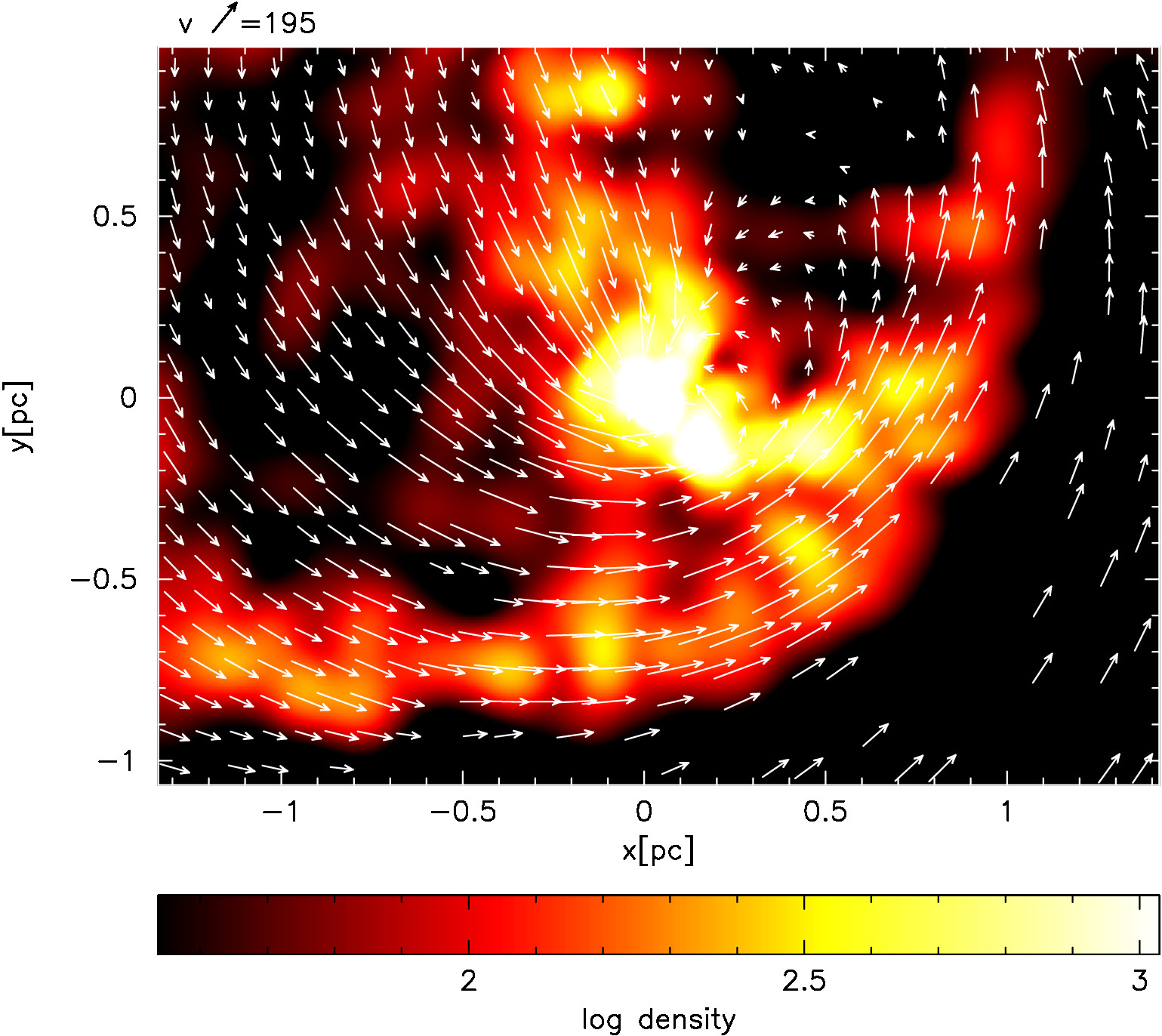}
\end{center}
\caption{Density cut in the z=0 pc plane in units of \density \ through the
inflow pattern of gas from a low-resolution test simulation that resembles the mini-spiral. The 2D velocity field agrees with observations 
presented in \protect\cite{2009ApJ...699..186Z}, but a detailed 3D comparison still has to be undertaken.
}
\label{spiral_1}
\end{figure}

The inflow pattern of gas especially at the beginning of our simulations always consists of multiple streams of
gas flowing into the central region from different points of the inner edge of the GD.
In one particular case of a low-resolution test simulation, the inflow pattern resembles the mini spiral surprisingly good.
We present this case in Fig. \ref{spiral_1}.
The mini spiral has already been credited as source for the black hole SgrA* in its bright state \citep{2012arXiv1211.1226C}.\\

A comparison to observations, e.g. Fig. 2 in \cite{2009ApJ...699..186Z} shows that the 2D velocity field matches observations quite well, 
especially the small loop forming at the upper right half of the plot. In our picture the eastern-arm (stream of gas on the left side of the plot)
crosses the black hole from below
and also forms the so called bar below the black hole. From the north we have the northern-arm flowing past the black hole on the left side,
colliding with the bar at the crossing point of the streams.\\

A detailed comparison of the 3D structure of the inflow pattern to the fit from observations by \cite{2009ApJ...699..186Z} is still necessary and 
will be postponed to a future publication. The formation of a structure similar to the mini-spiral seems to be a natural feature of our model
for the formation of the two counter-rotating sub-parsec scale accretion disks,
suggesting that we observe a similar event today at the GC which happened 10 Myr ago and lead to the formation of the disks.


\section{Discussion and Summary}
\label{summary}

We presented a simple, plausible model for the formation of two counter-rotating accretion disks.
This model is for the first time able to explain the formation of the puzzling configuration of the two stellar disks which are
observed at the Galactic Center (GC) without relying on long chains of individual events.\\

So far it was assumed that a single cloud with a very small (sub-parsec) impact parameter gets captured and disrupted by the
black hole to form a single accretion disk. To form a second disk, one would need within a short time a second, counter-rotating cloud
with equally small impact parameter.
Already the existence of a single cloud with a sub-parsec impact parameter is highly unlikely. It would have to be 
created very close to the black hole from the beginning.
However, the only real source of gas close to the black hole is the circum-nuclear disk (CND) (for the current event) or
our simulated gaseous disk (GD) (for the previous event),
so it is likely that the gas which formed the accretion disks originates there.\\

The CND/GD is placed far enough from the black hole so that collisions with clouds become possible. Observations even confirm that 
there is currently a cloud colliding with the CND.
Test simulations have shown that it is easy to create multiple accretion disks using this model. There are cases with up to four compact accretion disks forming.
Sometimes all the disks have the same rotational direction, sometimes only a single disk forms. But in all cases 
the formation of an accretion disk due to low angular momentum gas winding up around the black hole is possible.
The inflow patterns of the streams of gas can even resemble the mini-spiral which is currently observed at the GC.\\

Due to the high degree of freedom in our initial conditions we don't expect to get a perfect match to observations. 
For our high resolution simulation we took the parameters from a low resolution simulation that produced two prominent,
counter-rotating disks of roughly the correct size compared to observations, in order to show the basic mechanism of our proposed model.\\

The stellar disks themselves had a long time to evolve since their formation and there is a number of papers \citep{2009MNRAS.398..535U, 2011MNRAS.412..187K, 2009ApJ...697L..44M, 2012arXiv1210.4750U}
that deal with how the disks could become warped with time or increase their inclination.
In this work we concentrate on the initial formation of the progenitor accretion disks and hence a perfect fit to all observed 
characteristics of the stellar disks is not intended.
Test simulations have shown that increasing the mass imbalance above and below the xy-plane, as well as allowing a small cloud
offset in z-direction also leads to a larger inclination, thus an improved simulation later on could also be able to explain the high
inclination directly.\\

We have made a number of simplifications which could impact our results, which will be briefly presented here.
First we always assume the cloud and the GD to have equal mass. A naive improvement judging from Fig. \ref{toomre_1} 
would be to increase the mass in the GD so that the second high mass accretion disk becomes more massive, since the main contribution
to mass in this disk comes from the GD. This could also push the first high mass accretion disk over the fragmentation point.
In addition this would bring the disks closer to the observed mass ratio of 1:2 (10$^4$ \msol\  in the clockwise disk and
0.5 $\times$ 10$^4$ \msol\  in the counter-clockwise disk).\\

Feedback from the black hole would depend on the orientation of the black hole spin axis with respect to the accretion disks. 
A real first hint on the spin axis orientation might be given by the recent possible detection of a jet-like outflow from SgrA*
\citep{2012ApJ...758L..11Y}. Such a jet could prevent the formation of a second disk after material from the first disk feeds the black hole.\\

Another important point to keep in mind is the possible impact of stellar feedback. We do not include stellar feedback in our simulations.
If there are stars forming in the first accretion disk, they could possibly influence the formation of the second accretion disk, given enough 
time between the two events. Thus it is desirable that the disks form as short after one another as possible.\\

On the other hand, the disks must have enough time to fragment at all. This last point does not seem to be overall problematic judging from
our test simulations. Simulations of fragmenting high mass (10$^5$ \msol) accretion disks close ($<$ 0.5 pc) to SgrA* typically only
need around 0.2 to 0.3 Myr to form clumps and protostellar cores, as already pointed out in section \ref{result_mass}. Even in cases where we had four
high mass accretion disks forming, each single accretion disk existed for around 0.2 Myr and thus would have had enough time to fragment.\\

More problematic is the time between the formation of the two disks. In our case it takes roughly 1 Myr between formation of the two disks,
during which feedback from the stars forming in the first disk could become already important. 
However, taking the stellar models of \cite{2012A&A...537A.146E} into account this should be short enough, so that stellar feedback
even from high mass stars ($>$ 60 \msol) should not play a role.\\

Compared to observations we are well within the uncertainty of age determinations of the stars of 2 Myr.
The stellar and black hole feedback problem also applies to the low angular momentum cloud capture model and is not a special property of our particular
formation model.\\

Overall, seeing how easy it is to create multiple disks, we are still confident that our results will hold when including additional
physics. This first study already produced a reasonable fit when compared to observations and future refinements will surely 
increase the agreement. For the first time we are able to create two massive, sub-parsec scale counter-rotating accretion disks which are inclined
with respect to each other and with respect to the CND.


\section*{Acknowledgments}
We would like to thank the Reviewer for very useful comments and constructive criticism, that helped us to improve the paper.
This research was supported by the DFG cluster of excellence "Origin and Structure of the Universe"
and by the Deutsche Forschungsgemeinschaft priority program 1573 ("Physics of the Interstellar Medium").
Computer resources for this project have been provided by the Gauss Center for Supercomputing/Leibniz Supercomputing Center under grant: h0075.
Most of the plots have been created using the publicly available SPH visualization tool SPLASH by D.J. Price \citep{2007PASA...24..159P}.

\bibliographystyle{apj}
\bibliography{ref}


\label{lastpage}
\end{document}